\documentclass{article}
\usepackage{amssymb}
\usepackage{amsfonts}
\usepackage{amsmath}

\input{tcilatex}
\begin{document}

\begin{center}
\bigskip

{\large Hadronic Sector in the 4-d Pseudo-Conformal Field Theory}

{\large \ }\bigskip

C. N. Ragiadakos

email: ragiadak@gmail.com

\bigskip

\textbf{ABSTRACT}
\end{center}

The pseudo-conformal field theory (PCFT) is a 4-d action, which depends on
the lorentzian Cauchy-Riemann (LCR) structure. Like the 2-d linearized
string action, it does not depend on the metric tensor. But the invariance
under the pseudo-conformal transformations (in the terminology of E. Cartan
and Tanaka) imposes in the action the existence of a gauge field instead of
the scalar field of the string action. The tetrad of the LCR-structure
defines a class of metrics and a corresponding class of self dual 2-forms.
Soliton and multisoliton point of view of PCFT is described and related to
the Einstein derivation of the equations of motion. After the expansion of
the action around the static LCR-structure soliton, the quadratic part of
the Yang-Mills-like term implies a linear partial differential equation
(PDE). I solve this PDE using the Teukolsky method for the solution of the
electromagnetic field in the background of the Kerr black hole. The angular
and radial ODEs are different from the corresponding Teukolsky master
equations. The exact gauge field PDEs are also solved, using the fundamental
property of LCR-structure coordinates. The found solutions have colored
sources, which could be identified with the quarks.

\newpage

{\LARGE Contents}

\textbf{1. INTRODUCTION}

\textbf{2. SOLITONS AND MULTISOLITONS}

\textbf{3. SOLUTION\ OF\ THE\ STABILITY\ EQUATION}

\qquad 3.1 Study of the angular differential equation

\qquad 3.2 Study of the radial differential equation

\textbf{4. COLORED\ SOLITONS}

\qquad 4.1 A non-abelian solution

\textbf{5. PERSPECTIVES}

\qquad \newpage

\bigskip

\renewcommand{\theequation}{\arabic{section}.\arabic{equation}}

\section{INTRODUCTION}

\setcounter{equation}{0}

A renormalizable generally covariant 4-dimensional quantum field theory\cite%
{RAG1990}, which depends on the lorentzian Cauchy-Riemann (LCR) structure
and not the spacetime metric, seems to provide the appropriate framework to
describe current phenomenology. In my last work\cite{RAG2018a}, I derived
the electromagnetic and weak interactions of the standard model (SM), where
the particle-like static LCR-manifold is identified with the electron and
its complex conjugate LCR-manifold with the positron. The massless limit of
this LCR-manifold has only a left-handed part and it is identified with the
neutrino. The SM is the effective action implied after the introduction of
quantum fields for the electron and the neutrino solitons, in complete
analogy to the condensed matter considerations. The interaction terms are
introduced using the Bogoliubov-Medvedev-Polivanov (BMP)\cite{BOG1982}
S-matrix computational procedure, which in the present case happens to close
up to the well known SM action. The spontaneously broken $SU(2)\times U(1)$
group naturally emerges from the BMP closing up procedure. This internal
symmetry is purely effective, and implied by the fact that the initial
LCR-structure solitons (the massive electron and the massless neutrino)
break the symmetry. The renormalizability condition restricts the masses and
the coupling constants to take the appropriate relations. The same condition
does not permit the incorporation of the linearized soliton-graviton
interaction into the SM, because the closing up procedure generates
non-renormalizable terms. The gluonic Yang-Mills-like term in the present
action does not coincide with conventional quantum chromodynamics (QCD). In
the linearized gravity approximation, its Poincar\'{e} invariance is not
conventional. Besides we cannot introduce external sources in the action,
because its extended local symmetry and subsequently its formal
renormalizability will be destroyed. Therefore it cannot be included in the
effective SM quantum field theory\cite{BOG1975}. On the other hand ordinary
QCD has not yet provided a self-consistent mechanism for confinement. I
point out that the present gluon perturbative potential is linear, while QCD
implies the ($1/r$) Coulomb-like potential. In the present work I use
solitonic techniques\cite{RAJ} in the gauge field sector of the lagrangian.
I solve the soliton implied linearized stability partial differential
equations (PDE) and the exact (non-linear) gauge field PDEs in the static
LCR-manifold background.

Let me first present you the original idea\cite{RAG1990}$^{,}$\cite{RAG2008b}
to study LCR-structure dependent field theories. They emerged from the
observation that the linearized string action 
\begin{equation}
\begin{array}{l}
I_{S}=\frac{1}{2}\int d^{2}\!\xi \ \sqrt{-\gamma }\ \gamma ^{\alpha \beta }\
\partial _{\alpha }X^{\mu }\partial _{\beta }X^{\nu }\eta _{\mu \nu } \\ 
\end{array}
\label{i1}
\end{equation}%
does not essentially depend on the metric $\gamma ^{\alpha \beta }$ of the
2-dimensional surface, because in the light-cone coordinates $(\xi _{-},\
\xi _{+})$ the metric and the action take the form 
\begin{equation}
\begin{array}{l}
ds^{2}=2\gamma d\xi _{+}d\xi \\ 
\\ 
I_{S}=\int d^{2}\!z\ \partial _{-}X^{\mu }\partial _{+}X^{\nu }\eta _{\mu
\nu } \\ 
\end{array}
\label{i2}
\end{equation}%
We see that in these coordinates the action does not depend on the metric,
while it is not topological. This metric independence, without being
topological, is the crucial property of the action, which should be
transferred to four dimensions and not the simple Weyl invariance. That is,
the four dimensional analogous symmetry has to be a form of pseudo-conformal
symmetry (Cauchy-Riemann structure) and not the conventional Weyl symmetry.
The pioneers of the CR-structure, E. Cartan and Tanaka, used to call it
pseudo-conformal structure, therefore I prefer to call the present
lagrangian model pseudo-conformal field theory (PCFT) in order to point out
that it is essentially a 4-dimensional analogue of the 2-dimensional
conformal models.

Four dimensional spacetime metrics cannot generally take the form (\ref{i2}%
). Only metrics which admit two geodetic and shear free null congruences $%
\ell ^{\mu }\partial _{\mu },\ n^{\mu }\partial _{\mu }$ can take\cite%
{FLAHE1974}$^{,}$\cite{FLAHE1976} the analogous form 
\begin{equation}
\begin{array}{l}
ds^{2}=2g_{a\widetilde{\beta }}dz^{\alpha }dz^{\widetilde{\beta }}\quad
,\quad \alpha ,\widetilde{\beta }=0,1 \\ 
\end{array}
\label{i3}
\end{equation}%
where $z^{b}=(z^{\alpha }(x),z^{\widetilde{\beta }}(x))$ are generally
complex coordinates. The 4-dimensional analog to the Polyakov action (\ref%
{i2}) is the following metric independent Yang-Mills-like integral 
\begin{equation}
\begin{array}{l}
I_{G}=\int d^{4}\!z\ \sqrt{-g}g^{\alpha \widetilde{\alpha }}g^{\beta 
\widetilde{\beta }}F_{\!j\alpha \beta }F_{\!j\widetilde{\alpha }\widetilde{%
\beta }}=\int d^{4}\!z\ F_{\!j01}F_{\!j\widetilde{0}\widetilde{1}} \\ 
\\ 
F_{j_{ab}}=\partial _{a}A_{jb}-\partial _{a}A_{jb}-\gamma
\,f_{jik}A_{ia}A_{kb}%
\end{array}
\label{i4}
\end{equation}%
which depends on the LCR-structure coordinates, it does not depend on the
metric, but it is complex. We will finally assume as action either its real
or its imaginary part.

Notice the similarity of this 4-dimensional action with the 2-dimensional
action (\ref{i2}). In the place of the "field" $X^{\mu }$, which is
interpreted as the background 26-dimensional Minkowski spacetime in string
theory, we now have a gauge field $A_{j\nu }$, which we have to interpret as
the gluon, because the field equations generate a linear potential instead
of the Coulomb-like ($\frac{1}{r}$) potential of ordinary Yang-Mills action.

The present action is based on the lorentzian CR-structure\cite{RAG2013b},
which is determined by two real and one complex independent 1-forms ($\ell
,n,m,\overline{m}$). This LCR-structure tetrad satisfy the relations 
\begin{equation}
\begin{array}{l}
d\ell =Z_{1}\wedge \ell +i\Phi _{1}m\wedge \overline{m} \\ 
\\ 
dn=Z_{2}\wedge n+i\Phi _{2}m\wedge \overline{m} \\ 
\\ 
dm=Z_{3}\wedge m+\Phi _{3}\ell \wedge n \\ 
\end{array}
\label{i5}
\end{equation}%
where the vector fields $Z_{1\mu }\ ,\ Z_{2\mu }\ $ are real, the vector
field$\ Z_{3\mu }$ is complex, the scalar fields $\Phi _{1}\ ,\ \Phi _{2}$
are real and the scalar field$\ \Phi _{3}$ is complex. In the present model,
this structure essentially replaces the riemannian structure of the
spacetime in the Einstein general relativity. The form (\ref{i5}) is
completely integrable via the (holomorphic) Frobenius theorem, which implies
that the lorentzian CR-manifold (LCR-manifold) is defined\cite{BAOU} as a
4-dimensional real submanifold of $%
%TCIMACRO{\U{2102} }%
%BeginExpansion
\mathbb{C}
%EndExpansion
^{4}$ determined by four special (real) functions, 
\begin{equation}
\begin{array}{l}
\rho _{11}(\overline{z^{\alpha }},z^{\alpha })=0\quad ,\quad \rho
_{12}\left( \overline{z^{\alpha }},z^{\widetilde{\alpha }}\right) =0\quad
,\quad \rho _{22}\left( \overline{z^{\widetilde{\alpha }}},z^{\widetilde{%
\alpha }}\right) =0 \\ 
\end{array}
\label{i6}
\end{equation}%
where $\rho _{11}\ ,\ \rho _{22}$ are real and $\rho _{12}$ is a complex
function and $z^{b}=(z^{\alpha },z^{\widetilde{\alpha }}),\ \alpha =0,1$ are
the local structure coordinates in $%
%TCIMACRO{\U{2102} }%
%BeginExpansion
\mathbb{C}
%EndExpansion
^{4}$. Notice the special dependence of the defining functions on the
structure coordinates. They are not general functions of $z^{b}$. The
LCR-structure may be viewed as a restricted totally real CR-structure\cite%
{BAOU}. The separation of chiralities in the standard model is caused by
this property. The LCR-structure is more general than the riemannian
structure of general relativity and permits the invariance of the set of
solutions to the pseudo-conformal transformations\cite{RAG2013b}.

Using the properties of the structure coordinates we find 
\begin{equation}
\begin{array}{l}
F_{i01}F_{i\widetilde{0}\widetilde{1}}dz^{0}\wedge dz^{1}\wedge dz^{%
\widetilde{0}}\wedge dz^{\widetilde{1}}=\ell \wedge m\wedge n\wedge 
\overline{m}(\ell ^{\mu }m^{\nu }F_{i\mu \nu })(n^{\rho }\overline{m}%
^{\sigma }F_{i\rho \sigma }) \\ 
\\ 
\ell \wedge m\wedge n\wedge \overline{m}=d^{4}\!x\sqrt{-g}i \\ 
g=\det (g_{\mu \nu })=\det (\eta _{ab})[\det (e_{\mu }^{a})]^{2}=[\det
(e_{\mu }^{a})]^{2} \\ 
\end{array}
\label{i6a}
\end{equation}%
Hence taking the real part of (\ref{i4}) we find the following generally
covariant form 
\begin{equation}
\begin{array}{l}
I_{R}=\tint d^{4}\!x\sqrt{-g}i\{(\ell ^{\mu }m^{\nu }F_{i\mu \nu })(n^{\rho }%
\overline{m}^{\sigma }F_{i\rho \sigma })-(\ell ^{\mu }\overline{m}^{\nu
}F_{i\mu \nu })(n^{\rho }m^{\sigma }F_{i\rho \sigma })\} \\ 
\\ 
F_{j\mu \nu }=\partial _{\mu }A_{j\nu }-\partial _{\nu }A_{j\mu }-\gamma
\,f_{jik}A_{i\mu }A_{k\nu }%
\end{array}
\label{i7a}
\end{equation}%
If we take the imaginary part of (\ref{i4}) we find the following covariant
form 
\begin{equation}
\begin{array}{l}
I_{I}=\int d^{4}\!x\ \sqrt{-g}\ \left\{ \left( \ell ^{\mu }m^{\rho
}F_{\!j\mu \rho }\right) \left( n^{\nu }\overline{m}^{\sigma }F_{\!j\nu
\sigma }\right) +\left( \ell ^{\mu }\overline{m}^{\rho }F_{\!j\mu \rho
}\right) \left( n^{\nu }m^{\sigma }F_{\!j\nu \sigma }\right) \right\} \\ 
\end{array}
\label{i7}
\end{equation}%
In principle both actions are compatible with the LCR-struture. In fact the
gauge field strengths $F_{i\mu \nu }$ of these two actions are the each
other dual, because $\ell ^{\lbrack \mu }m^{\rho ]}$ and $n^{[\nu }\overline{%
m}^{\sigma ]}$ are self-dual (relative to their corresponding metric).

We also have to consider the additional action term with the integrability
conditions on the tetrad 
\begin{equation}
\begin{array}{l}
I_{C}=\int d^{4}\!x\ \sqrt{-g}\{\phi _{0}(\ell ^{\mu }m^{\nu }-\ell ^{\nu
}m^{\mu })(\partial _{\mu }\ell _{\nu })+ \\ 
\\ 
\qquad +\phi _{1}(\ell ^{\mu }m^{\nu }-\ell ^{\nu }m^{\mu })(\partial _{\mu
}m_{\nu })+\phi _{\widetilde{0}}(n^{\mu }\overline{m}^{\nu }-n^{\nu }%
\overline{m}^{\mu })(\partial _{\mu }n_{\nu })+ \\ 
\\ 
\qquad +\phi _{\widetilde{1}}(n^{\mu }\overline{m}^{\nu }-n^{\nu }\overline{m%
}^{\mu })(\partial _{\mu }\overline{m}_{\nu })+c.conj.\}%
\end{array}
\label{i8}
\end{equation}%
These Lagrange multipliers introduce the integrability conditions of the
tetrad and make the complete action $I=I_{G}+I_{C}$ self-consistent and the
usual quantization techniques may be used\cite{RAG1992}. The action is
formally renormalizable\cite{RAG2008a}, because it is dimensionless and
metric independent. Its path-integral quantization is also formulated\cite%
{RAG2017} as functional summation of open and closed 4-dimensional
lorentzian CR-manifolds in complete analogy to the summation of
2-dimensional surfaces in string theory\cite{POL}. These transition
amplitudes of a quantum theory of LCR-manifolds provides the self-consistent
algorithms for the computation of the physical quantities.

The LCR-structure defining tetrad is invariant under the following
tetrad-Weyl transformations

\begin{equation}
\begin{tabular}{l}
$\ell _{\mu }^{\prime }=\Lambda \ell _{\mu }\quad ,\quad n_{\mu }^{\prime
}=Nn_{\mu }\quad ,\quad m_{\mu }^{\prime }=Mm_{\mu }$ \\ 
\\ 
$\ell ^{\prime \mu }=\frac{1}{N}\ell ^{\mu }\quad ,\quad n^{\prime \mu }=%
\frac{1}{\Lambda }n^{\mu }\quad ,\quad m^{\prime \mu }=\frac{1}{\overline{M}}%
m^{\mu }$ \\ 
\end{tabular}
\label{i9}
\end{equation}%
with non-vanishing $\Lambda \ ,\ N\ ,\ M$. I point out that we have not yet
introduced a metric. The tetrad with upper and lower indices is simply a
basis of tangent and cotangent spaces. But the tetrad does define a class $%
[g_{\mu \nu }]$ of symmetric tensors \ 
\begin{equation}
\begin{array}{l}
g_{\mu \nu }=\ell _{\mu }n_{\nu }+\ell _{\nu }n_{\mu }-m_{\mu }\overline{m}%
_{\nu }-m_{\nu }\overline{m}_{\mu } \\ 
\end{array}
\label{i10}
\end{equation}%
Every such tensor may be used as a metric to build up the riemannian
geometry of general relativity, because its local signature is ($1,-1,-1,-1$%
). But this form always admits two geodetic and shear-free null congruences
and hence it does not cover all the metrics of general relativity. The
existence of the metric permits us to consider solitonic LCR-manifolds with
energy-momentum and angular momentum in the linearized Einstein general
relativity, without requiring any finite integral on regular configurations
(ordinary solitonic postulate). The purpose of the section II of the present
work is to clarify the interacting solitonic LCR-manifolds.

The defining relations (\ref{i6}) of the quite general class of LCR-manifolds%
\cite{RAG2013b} take the following form of real surfaces of the grassmannian
manifold $G_{4,2}$ 
\begin{equation}
\begin{array}{l}
\rho _{11}(\overline{X^{m1}},X^{n1})=0=\rho _{22}(\overline{X^{m2}},X^{n2})
\\ 
\rho _{12}(\overline{X^{m1}},X^{n2})=0 \\ 
K(X^{mj})=0%
\end{array}
\label{i11}
\end{equation}%
where all the functions are homogeneous relative to the homogeneous
coordinates $X^{n1}$ and $X^{n2}$\ independently, which must be roots of the
homogeneous holomorphic Kerr polynomial $K(Z^{m})$. The charts of its
typical non-homogeneous coordinates are determined by the invertible pairs
of rows. If the first two rows constitute an invertible matrix, the chart is
determined by $\det \lambda \neq 0$ and the corresponding affine space
coordinates $r$ are defined by 
\begin{equation}
\begin{array}{l}
X=%
\begin{pmatrix}
X^{01} & X^{02} \\ 
X^{11} & X^{12} \\ 
X^{21} & X^{22} \\ 
X^{31} & X^{32}%
\end{pmatrix}%
=\left( 
\begin{array}{c}
\lambda ^{Aj} \\ 
-ir_{A^{\prime }A}\lambda ^{Aj}%
\end{array}%
\right) \\ 
\\ 
r_{A^{\prime }A}=\eta _{ab}r^{a}\sigma _{A^{\prime }A}^{b} \\ 
\end{array}
\label{i12}
\end{equation}

In this context, we see that the LCR-structures determined by the relations 
\begin{equation}
\begin{array}{l}
\overline{X^{mi}}E_{mn}X^{nj}=0\quad ,\quad K(X^{mj})=0 \\ 
\\ 
E=%
\begin{pmatrix}
0 & I \\ 
I & 0%
\end{pmatrix}
\\ 
\end{array}
\label{i13}
\end{equation}%
are flat, i.e. they generate a minkowiskian class of metrics $[\eta _{\mu
\nu }]$. Notice that $SU(2,2)$ is the symmetry group of these solutions. Its
Poincar\'{e} subgroup is identified with the observed Poincar\'{e} symmetry
in nature.

It is more convenient to give the general solution (\ref{i11}) in the
physically interesting form 
\begin{equation}
\begin{array}{l}
\overline{X^{mi}}E_{mn}X^{nj}=G_{ij}(\overline{X^{mi}},X^{mj})\quad ,\quad
K(X^{mj})=0 \\ 
\end{array}
\label{i14}
\end{equation}%
where the $2\times 2$ matrix with the precise dependence on the left and
right columns of the homogeneous coordinates essentially generates gravity.
The general mathematical context of these algebraic equations is the
following: The Kerr polynomial $K(Z^{m})$ determines a hypersurface of $%
CP^{3}$. The irreducible quadratic polynomial implies the electron and the
reducible one yields the neutrino. The parametric representation\cite{GRIF}
of the canonical form of the irreducible quadratic representation is \ 
\begin{equation}
\begin{array}{l}
Z^{m}=(t_{1}^{2}+t_{2}^{2}+1):i(t_{1}^{2}+t_{2}^{2}-1):-2it_{1}:-2it_{2} \\ 
\end{array}
\label{i15}
\end{equation}%
This is essentially equivalent to the introduction of the Newman complex
trajectory, that I have extensively used in order to generate our intuition.
The LCR-structure is determined by two distinct points $X^{n1}(z^{\alpha })$
and $X^{n2}(z^{\widetilde{\alpha }})$ which belong to the two sheets of the
quadratic hypersurface. At the finite branch point, the line (determined by
the two points) becomes tangent to the surface and the LCR-structure has a
"singularity".

Using the results of the E. Cartan work on the automorphisms of the
3-dimensional CR-structures, I found\cite{RAG2017} the automorphisms of the
LCR-structure. The case of two commuting generators coincides with the two
commuting generators of the Poincar\'{e} group. This permits the computation%
\cite{RAG2008b} of a static massive soliton and its massless stationary
limit, as the basic free solitons of the model, which are identified with
the electron and the neutrino of the SM. The stability of these solitons is
assured by their topological characteristics and the LCR-structure relative
invariants. The electron LCR-manifold is determined by an irreducible
quadratic polynomial (\ref{i11}) of $CP^{3}$, while the neutrino one is
determined by the corresponding reducible quadratic polynomial.

In section II, the classical configurations which describe the
electron-electron and the positronium bound state are revealed, in complete
analogy to sine-Gordon soliton-soliton scattering and the "breather". It is
explicitly proven that the compatibility of the LCR-structures is achieved
if the trajectories of the electron LCR-structure solitons satisfy the
equations of motion. In the case of the electron-positron system the
positronium is expected.

In section III, the linearized part of the gluon field partial differential
equations (PDE) are solved in the massive static soliton (electron)
background. Following the Teukolsky technique\cite{CHAND} of the solution of
the analogous problem of a photon in the Kerr black hole, I first achieve to
bring them to a form, which permits a separation of variables. The implied
angular and radial ordinary differential equations (ODE) are completely
different to the Teukolsky master equations.

In section IV, the exact non-abelian gauge field PDEs in the static
LCR-manifold background are solved. The solution has monopole sources, which
cannot be removed through a duality rotation, even in the linearized
approximation, because there is one gluon charge while the gauge field
components are determined by the dimension of Lie algebra. The reasonable
physical interpretation is to identify these colored monopole sources with
the quarks, which should be confined by the Dirac potential singularity.

\section{SOLITONS AND\ MULTISOLITONS}

\setcounter{equation}{0}

Ordinary solitons are field configurations with finite energy-momentum and
angular-momentum. A typical example of solitons are the finite energy
classical solutions of the 2-dimensional sine-Gordon equation\cite{RAJ}.
These objects are additional states (besides the ordinary mesons) in the
quantum Fock space of the sine-Gordon lagrangian. The general method to
study the quantum scattering of solitons is to start from the classical
scattering solution and proceed with the higher $\hbar $ terms. The bound
states of two quantum solitons is indicated by the existence of a
corresponding classical solution, with the typical example the sine-Gordon
"breather". This procedure is well known\cite{RAJ} and I am not going to
review it. Through the present work I will refer to this picture in order to
facilitate the reader to understand my solitonic approach in the context of
the present 4-dimensional PCFT. I will precisely briefly review\cite{RAG1999}
the massive static solution\cite{RAG1991} of PCFT. The required
energy-momentum and the angular momentum of the soliton is defined using the
ordinary linearized gravity definition\cite{MTW} with metric (\ref{i10})
defined by the LCR-structure. The emergence of the gravity permit us to
define the energy-momentum without needing the finite integrals condition of
ordinary solitons.

Newman has found\cite{NEWM1973}$^{,}$\cite{NEWM2004} that the Kerr function
condition (for a null congruence to be geodetic and shear-free) may be
replaced with a (generally complex) trajectory $\xi ^{a}(\tau )$. In the
context of conventional algebraic geometry, the Newman complex trajectory is
a parametrization of the physically interesting hypersurfaces of $CP^{3}$.
For the case of the quadric, the linear trajectory is easily derived from (%
\ref{i15}). In the present case of the LCR-structure formalism, this is done
by assuming that the $G_{4,2}$ two homogeneous coordinates $i=1,2$ have the
form \ 
\begin{equation}
\begin{array}{l}
X^{i}=%
\begin{pmatrix}
\lambda ^{i} \\ 
-i\xi (\tau _{i})\lambda ^{i}%
\end{pmatrix}
\\ 
\\ 
\xi (\tau _{i})=\xi ^{a}(\tau _{i})\sigma ^{b}\eta _{ab} \\ 
\end{array}
\label{s1}
\end{equation}%
where $\sigma ^{b}$and $\eta _{ab}$ are the Pauli matrices and the Minkowski
metric respectively. Here, I have to point out that the consideration of two
generally different complex Kerr homogeneous functions is somehow
misleading. In conventional algebraic geometry, the notion of reducible
polynomial is used. The irreducible quadratic Kerr polynomial (\ref{i11}) of
the electron LCR-structure corresponds to the complex trajectory $\xi
^{a}=(\tau ,0,0,ia)$.

The flatprint LCR-structure coordinates are determined by the condition \ 
\begin{equation}
\begin{array}{l}
(x-\xi (\tau _{i}))\lambda ^{i}=0 \\ 
\end{array}
\label{s2}
\end{equation}%
that admits one non-vanishing solution for every column $i=1,2$ of the
homogeneous coordinates of $G_{4,2}$. This is possible if \ 
\begin{equation}
\begin{array}{l}
\det (x-\xi (\tau _{i}))=\eta _{ab}(x^{a}-\xi ^{a}(\tau _{i}))(x^{b}-\xi
^{b}(\tau _{i}))=0 \\ 
\end{array}
\label{s3}
\end{equation}%
which gives the two solutions $z^{0}=\tau _{1}(x)$ and $z^{\widetilde{0}%
}=\tau _{2}(x)$. The other structure coordinates are $z^{1}=\frac{\lambda
^{11}}{\lambda ^{01}}$ and $z^{\widetilde{1}}=-\frac{\lambda ^{02}}{\lambda
^{12}}$ where 
\begin{equation}
\begin{array}{l}
\lambda ^{Aj}=%
\begin{pmatrix}
(x^{1}-ix^{2})-(\xi ^{1}(\tau _{j})-i\xi ^{2}(\tau _{j})) \\ 
(x^{0}-x^{3})-(\xi ^{0}(\tau _{j})-\xi ^{3}(\tau _{j}))%
\end{pmatrix}
\\ 
\end{array}
\label{s4}
\end{equation}%
Notice that the trajectory technique for computation of the structure
coordinates incorporates the notion of the classical causality, which is
apparently respected by (\ref{s3}).

The singularity of the flatprint LCR-structure occurs at $\det [\lambda
^{A1}(x),\lambda ^{B2}(x)]=0$. Recall that the left and right columns of the
homogeneous coordinates of $G_{4,2}$ may be determined ("move") with
different trajectories, if the corresponding homogeneous Kerr polynomial is
reducible. In the simple case when both move with the same trajectory $\xi
^{a}(\tau )=(\tau ,\xi ^{1}(\tau ),\xi ^{2}(\tau ),\xi ^{3}(\tau ))$, the
singularity occurs at $\tau _{1}(x)=\tau _{2}(x)$, which yields 
\begin{equation}
\begin{array}{l}
(x^{i}-\xi ^{i}(t))(x^{j}-\xi ^{j}(t))\delta _{ij}=0 \\ 
\end{array}
\label{s5}
\end{equation}%
If $\xi _{R}^{i}$ and $\xi _{I}^{i}$ are the real and imaginary parts of the
trajectory, we find that the locus of the solitonic LCR-structure is 
\begin{equation}
\begin{array}{l}
(x^{i}-\xi _{R}^{i}(t))(x^{j}-\xi _{R}^{j}(t))\delta _{ij}-\xi
_{I}^{i}(t)\xi _{I}^{j}(t)\delta _{ij}=0 \\ 
\\ 
(x^{i}-\xi _{R}^{i}(t)\xi _{I}^{j}(t)\delta _{ij}=0 \\ 
\end{array}
\label{s6}
\end{equation}%
Note that if $\xi _{I}^{j}(t)$ is bounded, the LCR-structure may be
interpreted as a soliton with trajectory $\xi _{R}^{i}(t)$ and a locus at
(the perimeter of) the circle of radius $(\xi _{I}^{i}(t))^{2}$ around its
trajectory. This locus (a two dimensional surface) is a singularity of the
gravitational potential and a source of the corresponding gravitational
radiation, but it is not a singularity of the LCR-structure viewed as a
surface of the $G_{4,2}$ grassmannian, because the matrix $X^{mi}$ has not
rank two at this surface.

The (class of) symmetric tensors $g_{\mu \nu }$ are identified with the
Einstein metric, which defines the curvature and the Einstein tensor $E^{\mu
\nu }$. Because of the singularity of the metric, the Einstein tensor must
be singular at the cylinder (\ref{s6}) of $%
%TCIMACRO{\U{211d} }%
%BeginExpansion
\mathbb{R}
%EndExpansion
^{4}$ and the circumference of $%
%TCIMACRO{\U{211d} }%
%BeginExpansion
\mathbb{R}
%EndExpansion
^{3}$. Besides the metric tensor, the LCR-structure tetrad defines the
(class of) self-dual 2-forms

\begin{equation}
\begin{array}{l}
V=2\ell \wedge n-2m\wedge \overline{m} \\ 
\end{array}
\label{s7}
\end{equation}%
Notice that this 2-form and the metric define the complex tensor

\begin{equation}
\begin{array}{l}
V_{\nu }^{\mu }=\ell ^{\mu }n_{\nu }-n^{\mu }\ell _{\nu }-m^{\mu }\overline{m%
}_{\nu }+\overline{m}^{\mu }m_{\nu } \\ 
\end{array}
\label{s8}
\end{equation}%
which is invariant under the tetrad-Weyl transformation. The LCR-structure
tetrad vectors\ are eigenvectors of this tensor. Hence this tensor
completely defines the LCR-structure. It is the complex pseudo-hermitian
tensor, introduced by Flaherty\cite{FLAHE1974}$^{,}$\cite{FLAHE1976} to
study the metrics with two geodetic and shear free congruences.

Through out the present work I will use the following form of the static
LCR-structure%
\begin{equation}
\begin{array}{l}
\ell _{\mu }dx^{\mu }=dt-\frac{\rho ^{2}}{\Delta }dr-a\sin ^{2}\theta
d\varphi \\ 
n_{\mu }dx^{\mu }=\frac{\Delta }{2\rho ^{2}}(dt+\frac{\rho ^{2}}{\Delta }%
dr-a\sin ^{2}\theta d\varphi ) \\ 
m_{\mu }dx^{\mu }=\frac{1}{\eta \sqrt{2}}(ia\sin \theta dt-\rho ^{2}d\theta
-i(r^{2}+a^{2})\sin \theta d\varphi ) \\ 
\\ 
\eta =r+ia\cos \theta \quad ,\quad \rho ^{2}=\eta \overline{\eta }\quad
,\quad \sqrt{-g}=\rho ^{2}\sin \theta \\ 
\Delta =r^{2}-2Mr+a^{2}+q^{2}%
\end{array}
\label{s9}
\end{equation}%
Its contravarient components are 
\begin{equation}
\begin{array}{l}
\ell ^{\mu }\partial _{\mu }=\frac{1}{\Delta }((r^{2}+a^{2})\partial
_{t}+\Delta \partial _{r}+a\partial _{\varphi }) \\ 
n^{\mu }\partial _{\mu }=\frac{1}{2\rho ^{2}}((r^{2}+a^{2})\partial
_{t}-\Delta \partial _{r}+a\partial _{\varphi }) \\ 
m^{\mu }\partial _{\mu }=\frac{1}{\eta \sqrt{2}}(ia\sin \theta \partial
_{t}+\partial _{\theta }+\frac{i}{\sin \theta }\partial _{\varphi }) \\ 
\end{array}
\label{s10}
\end{equation}
We will also need its spin coefficients%
\begin{equation}
\begin{array}{l}
\varepsilon =0\quad ,\quad \beta =\frac{\cos \theta }{\sin \theta \eta 2%
\sqrt{2}}\quad ,\quad \pi =\frac{ia\sin \theta }{(\overline{\eta })^{2}\sqrt{%
2}} \\ 
\tau =-\frac{ia\sin \theta }{\rho ^{2}\sqrt{2}}\quad ,\quad \rho =-\frac{1}{%
\overline{\eta }}\quad ,\quad \mu =-\frac{\Delta }{2\rho ^{2}\overline{\eta }%
} \\ 
\gamma =-\frac{\Delta }{2\rho ^{2}\overline{\eta }}+\frac{r-M}{2\rho ^{2}}%
\quad ,\quad \alpha =\pi -\overline{\beta }=\frac{ia\sin \theta }{(\overline{%
\eta })^{2}\sqrt{2}}-\frac{\cos \theta }{\sin \theta \overline{\eta }2\sqrt{2%
}}%
\end{array}
\label{s11}
\end{equation}

The tetrad-Weyl factors have been chosen such that to give the Kerr-Newman
manifold. They are imposed\cite{RAG1999} by the existence of the classical
electric charge and Poincar\'{e} charges, as follows

1) The self-dual 2-form (\ref{s7}) of the static soliton with the precise
tetrad-Weyl factors has the characteristic property to admit a complex
multiplication function such that%
\begin{equation}
\begin{array}{l}
F^{+}=\frac{C}{(r-ia\cos \theta )^{2}}(2\ell \wedge n-2m\wedge \overline{m})
\\ 
\\ 
dF^{+}=0 \\ 
\end{array}
\label{s12}
\end{equation}%
outside the singularity locus of the LCR-manifold, where $C$ is an arbitrary
complex constant. Hence for an arbitrary complex constant $C$, this complex
2-form defines a real 2-form $F$ such that%
\begin{equation}
\begin{array}{l}
dF=-\ast j_{m}\quad ,\quad d\ast F=-\ast j_{e} \\ 
\end{array}
\label{s12a}
\end{equation}%
where $j_{e}$ and $j_{m}$ are the "electric" and "magnetic" currents with
compact support. These are apparently analogous to the symmetric Maxwell
equations (with both electric and magnetic monopoles), which were used by
Dirac to prove the quantization of the electric charge. This implies that
the general electric charge is quantized\cite{RAG1999}. But the symmetry
under the duality rotation absorbs the magnetic charge (or electric charge)
leaving detectable only one kind of monopoles, as observed in
electrodynamics. That is, here we have a "self-quantization" of the electric
charge. But once fixed, the conserved electric charge reduces the general
tetrad-Weyl symmetry (\ref{i9}) down to the ordinary Weyl symmetry of the
electromagnetic field.

2) The precise tetrad-Weyl factors give a metric which coincides with its
linearized gravity approximation, and hence define the Poincar\'{e}
conserved quantities without any "approximation". This fact fixes the
remaining ordinary Weyl transformation.

The SM was derived by identifying the electron as the static LCR-manifold
determined by the complex linear trajectory%
\begin{equation}
\begin{array}{l}
\xi ^{b}(s)=v^{b}s+c^{b}+ia^{b}\quad ,\quad \overset{.}{(\xi ^{b}}%
)^{2}=(v^{b})^{2}=1 \\ 
\end{array}
\label{s13}
\end{equation}%
where $v^{b},c^{b},a^{b}$\ are the real constants, which represent the
constant velocity, the initial position and the spin of the classical
configuration of the electron. The closed self-dual 2-form (\ref{s12}) of
the static LCR-structure is identified with the self-dual 2-form $F^{+}$ of
the electromagnetic field. Hence the intuition suggests to consider the
particle-like LCR-manifold determined by a general complex trajectory, which
asymptotically at $t\rightarrow $ $\mp \infty $ describes an electron
interacting with an "external" field. In fact this is the electron-electron
current that was used to derive\cite{RAG2018a} the effective SM action.

The success of the leptonic standard model action is well known and its
derivation is a great success of the 4-d PCFT. It is an effective quantum
field theoretic model analogous to the ones used in condensed matter
physics. But the fundamental question "how can we compute the
electron-electron scattering in the context of PCFT?" remains. The abstract
answer to use the functional integral over the LCR-manifolds\cite{RAG2017}
is not a practical answer, because we do not know how to compute it. On the
other hand the "solitonic" picture permits us to use our computational
experience\cite{RAJ} with the 2-dimensional models, if we find the
multisolitons.

This point of view suggests to identify the classical solitonic
configuration of the electron-electron elastic scattering with the
LCR-manifold, which has two holes determined by non-intersecting general
complex trajectories $\xi _{1}^{b}$ and $\xi _{2}^{b}$, which become (time)
asymptotically linear. This can be used as a multisoliton, in complete
analogy to the calculations of the 2-dimensional solitonic models.

The Flaherty pseudo-hermitian structure must be uniquely defined outside the
"holes". That is, the class of metrics and the class of 2-forms must have
representatives uniquely defined outside the "holes" of the trajectories. In
the linearized gravity approximation%
\begin{equation}
\begin{array}{l}
g_{\mu \nu }\simeq \eta _{\mu \nu }+h_{\mu \nu }\quad ,\quad \widehat{h}%
_{\mu \nu }=h_{\mu \nu }-\frac{1}{2}\eta _{\mu \nu }h_{\rho }^{\rho }\quad
,\quad \partial _{\mu }\widehat{h}^{\mu \nu }=0 \\ 
\end{array}
\label{s14}
\end{equation}%
we have to solve the gravitational and electromagnetic laplacian problems%
\begin{equation}
\begin{array}{l}
\partial ^{2}\widehat{h}_{\mu \nu }=0\quad ,\quad dF^{+}=0 \\ 
\end{array}
\label{s15}
\end{equation}%
taking into account the existence of the two "holes" of the trajectories. It
is exactly the Einstein framework that implied the identification of the
"holes" trajectories with the geodetic trajectories. The self-consistency
conditions imposed to the momentum and the spin of every "hole" 
\begin{equation}
\begin{array}{l}
p_{j}^{\mu }=\dint\limits_{(j)}T^{\mu 0}d^{3}x\quad ,\quad s_{j}^{\mu \nu
}=\dint\limits_{(j)}(x^{\mu }T^{\nu 0}-x^{\nu }T^{\mu 0})d^{3}x \\ 
\end{array}
\label{s16}
\end{equation}%
to satisfy the equations of motion.

In the present case the trajectory $\zeta ^{i}(t,\varphi _{(k)})$ of every
point (fixed with the variable $\varphi _{(k)}$) of the moving ring has the
form 
\begin{equation}
\begin{array}{l}
\zeta ^{i}(t,\varphi _{(k)})=\xi _{R}^{i}(t)+\widehat{c}^{i}a(t)\cos \varphi
_{(k)}+\widehat{d}^{i}a(t)\sin \varphi _{(k)} \\ 
\\ 
\widehat{\xi _{I}}^{i}=\frac{\xi _{I}^{i}}{\sqrt{\xi _{I}^{2}}}\quad ,\quad
a^{2}(t)=\xi _{I}^{2}=\xi _{I}^{i}\xi _{I}^{i} \\ 
\end{array}
\label{s16a}
\end{equation}%
where $(\widehat{c}^{i},\widehat{d}^{i},\widehat{\xi _{I}}^{i})$\ is an
orthogonal local vector basis. Then the distributional currents of every
source-"hole" are 
\begin{equation}
\begin{array}{l}
j_{(k)}^{\mu }=\dint d\varphi _{(k)}q(\varphi _{(k)})\overset{.}{\zeta ^{\mu
}}(t,\varphi _{(k)})\delta (x^{i}-\zeta ^{i}(t,\varphi _{(k)}))\quad ,\quad
\zeta ^{\mu }(t,\varphi _{(k)})=(t,\zeta ^{i}(t,\varphi _{(k)})) \\ 
\\ 
T_{(k)}^{\mu \nu }=\dint d\varphi _{(k)}\mu (\varphi _{(k)})\overset{.}{%
\zeta ^{\mu }}(t,\varphi _{(k)})\overset{.}{\zeta ^{\nu }}(t,\varphi
_{(k)})\delta (x^{i}-\zeta ^{i}(t,\varphi _{(k)})) \\ 
\end{array}
\label{s16b}
\end{equation}%
where $q(\varphi _{(k)})$ and $\mu (\varphi _{(k)})$ are the charge and mass
densities on the ring. Recall that in the linearized gravity approximation,
the conservation of the energy-momentum current 
\begin{equation}
\begin{array}{l}
T_{(Gravity)}^{\mu \nu }=T_{(p)}^{\mu \nu }+T_{(EM)}^{\mu \nu }\quad ,\quad
\partial _{\mu }T_{(Gravity)}^{\mu \nu }=0 \\ 
\\ 
T_{(p)}^{\mu \nu }=\dsum\limits_{k}T_{(k)}^{\mu \nu }\quad ,\quad
T_{(EM)}^{\mu \nu }=-\frac{1}{4\pi }[F^{\mu \rho }F^{\nu \sigma }\eta _{\rho
\sigma }-\frac{1}{4}\eta ^{\mu \nu }F^{\rho \sigma }F_{\rho \sigma }] \\ 
\partial _{\mu }T_{(p)}^{\mu \nu }=\mu \frac{d^{2}\xi ^{\nu }}{dt^{2}}%
=F^{\nu \rho }j_{\rho }\quad ,\quad \partial _{\mu }T_{(EM)}^{\mu \nu
}=-F^{\nu \rho }j_{\rho }%
\end{array}
\label{s16c}
\end{equation}%
needs the use of the Lorentz force to be satisfied. In the hadronic sector
the analogous formalism is going to imply the "gluonic Lorentz force", which
in the context of PCFT is different from that of chromodynamics.

The derivations of the equations of motion, using either the Einstein\cite%
{EIH1938} or the Fock-Papapetrou\cite{PAP1948} points of view, have been
extensively studied and no review is needed. This procedure has to be
accommodated to include electromagnetic interaction. The Chase\cite{CHASE}
procedure may be used to find the equations of motion and the corresponding
metric and electromagnetic tensor at every order of the $\frac{1}{c}$
expansion. These two tensors define the Flaherty pseudo-hermitean tensor at
every order. Hence its integrability condition will assure the existence of
the LCR-structure at every order.

\section{SOLUTION\ OF\ THE\ STABILITY\ EQUATION}

\setcounter{equation}{0}

A characteristic ingredient of the solitonic technique is the investigation
of the linearized field equations in the soliton background. The
2-dimensional kink soliton of the $\phi ^{4}$ model\cite{RAJ} makes a bound
state with the perturbative meson. At the classical level this state appears
as a discrete energy level of the stability quadratic part, after the
expansion of the lagrangian around the $\phi _{kink}$ configuration. In the
case of astrophysics, perturbations are taken around the black hole solutions%
\cite{CHAND}. The yielded linearized PDEs are studied for (quasi-)normal
modes. Analogous investigations may be performed in the present PCFT model
after the expansion of the gluon field $A_{j\mu }$ action around the static
LCR-structure. The modes of this PDE are colored and apparently they have to
be related with hadronic phenomena. In this section I will solve these
partial differential equations. The mathematical problem is analogous to the
Teukolsky procedure for the solution of the photon stability PDE in the
Kerr-Newman black hole\cite{CHAND}, but apparently the equations are
different. The present PDE derived from the action $I_{I}$ (\ref{i7}) is 
\begin{equation}
\begin{array}{l}
\frac{1}{\sqrt{-g}}\partial _{\mu }\{\sqrt{-g}[(\ell ^{\mu }m^{\tau }-\ell
^{\tau }m^{\mu })(\phi _{j2})+(n^{\mu }\overline{m}^{\tau }-n^{\tau }%
\overline{m}^{\mu })(-\phi _{j0})+ \\ 
\qquad +(\ell ^{\mu }\overline{m}^{\tau }-\ell ^{\tau }\overline{m}^{\mu })(%
\overline{\phi _{j2}})+(n^{\mu }m^{\tau }-n^{\tau }m^{\mu })(\overline{-\phi
_{j0}})]\}=0 \\ 
\\ 
\phi _{j0}=-\ell ^{\mu }m^{\nu }(\partial _{\mu }A_{j\nu }-\partial _{\nu
}A_{j\mu }) \\ 
\phi _{j1}=-\frac{1}{2}(\ell ^{\mu }n^{\nu }+\overline{m}^{%
%TCIMACRO{\U{3bc} }%
%BeginExpansion
\mu
%EndExpansion
}m^{\nu })(\partial _{\mu }A_{j\nu }-\partial _{\nu }A_{j\mu }) \\ 
\phi _{j2}=n^{\mu }\overline{m}^{\nu }(\partial _{\mu }A_{j\nu }-\partial
_{\nu }A_{j\mu })%
\end{array}
\label{q1}
\end{equation}%
For convenience, I use the Newman-Penrose (NP) quantities $\phi _{0},\ \phi
_{1},\ \phi _{2}$ without the color index, because in the present linearized
terms the gluon interaction does not appear. The separation of variables is
achieved with the tetrad form (\ref{s10}) and the NP quantities (and not the
gluon potential) like in the already solved problem\cite{CHAND} of the
photon scattering with the Kerr black hole. Applying the tetrad form (\ref%
{s10}) and its spin coefficients (\ref{s11}), these PDEs take the final form%
\begin{equation}
\begin{array}{l}
\phi _{0}^{\prime }=\frac{\Delta }{2\overline{\eta }}\phi _{0}\quad ,\quad
\phi _{2}^{\prime }=\frac{\rho ^{2}}{\eta }\phi _{2} \\ 
m^{\prime \mu }\partial _{\mu }[\sin \theta \overline{\phi _{0}^{\prime }}]+%
\overline{m}^{\prime \mu }\partial _{\mu }[\sin \theta \phi _{0}^{\prime }]=0
\\ 
m^{\prime \mu }\partial _{\mu }[\sin \theta \phi _{2}^{\prime }]+\overline{m}%
^{\prime \mu }\partial _{\mu }[\sin \theta \overline{\phi _{2}^{\prime }}]=0
\\ 
-n^{\prime \mu }\partial _{\mu }[\sin \theta \phi _{0}^{\prime }]+\ell
^{\prime \mu }\partial _{\mu }[\sin \theta \overline{\phi _{2}^{\prime }}]=0
\\ 
\end{array}
\label{q2}
\end{equation}%
with the tetrad redefined as follows 
\begin{equation}
\begin{array}{l}
\ell ^{\prime \mu }\partial _{\mu }=\ell ^{\mu }\partial _{\mu }=\frac{%
r^{2}+a^{2}}{\Delta }\partial _{t}+\partial _{r}+\frac{a}{\Delta }\partial
_{\varphi } \\ 
n^{\prime \mu }\partial _{\mu }=\frac{2\rho ^{2}}{\Delta }n^{\mu }\partial
_{\mu }=\frac{r^{2}+a^{2}}{\Delta }\partial _{t}-\partial _{r}+\frac{a}{%
\Delta }\partial _{\varphi } \\ 
m^{\prime \mu }\partial _{\mu }=\eta \sqrt{2}m^{\mu }\partial _{\mu }=ia\sin
\theta \partial _{t}+\partial _{\theta }+\frac{i}{\sin \theta }\partial
_{\varphi } \\ 
\end{array}
\label{q3}
\end{equation}

Notice that these PDEs do not contain the NP field component $\phi _{1}$.
This component appears by imposing the integrability of gluon 2-form. Using
the following notation 
\begin{equation}
\begin{array}{l}
F=\frac{\phi _{1}}{2}V-\phi _{2}V^{0}+\phi _{0}V^{\widetilde{0}}+c.c. \\ 
\\ 
V^{0}=\ell \wedge m\ ,\ V^{\widetilde{0}}=n\wedge \overline{m}\ ,\ V=2\ell
\wedge n-2m\wedge \overline{m} \\ 
\end{array}
\label{q4}
\end{equation}%
and the formulas

\begin{equation}
\begin{array}{l}
dV^{0}=[(2\varepsilon -\rho )n+(\tau -2\beta )\overline{m}]\wedge V^{0} \\ 
dV^{\widetilde{0}}=[(\mu -2\gamma )\ell +(2\alpha -\pi )m]\wedge V^{%
\widetilde{0}} \\ 
dV=[2\mu \ell -2\rho n-2\pi m+2\tau \overline{m}]\wedge V \\ 
\end{array}
\label{q5}
\end{equation}%
the definition of the gluon field $F_{j\mu \nu }$ via the gluonic potential $%
A_{j\mu }$ implies the linearized integrability condition 
\begin{equation}
\begin{array}{l}
(\overline{m}+\pi -2\alpha )\phi _{0}-(m+\overline{\pi }-2\overline{\alpha })%
\overline{\phi _{0}}=(\ell -2\rho )\phi _{1}-(\ell -2\overline{\rho })%
\overline{\phi _{1}} \\ 
(m+2\beta -\tau )\phi _{2}-(\overline{m}+2\overline{\beta }-\overline{\tau })%
\overline{\phi _{2}}=(n+2%
%TCIMACRO{\U{3bc} }%
%BeginExpansion
\mu
%EndExpansion
)\phi _{1}-(n+2\overline{%
%TCIMACRO{\U{3bc} }%
%BeginExpansion
\mu
%EndExpansion
})\overline{\phi _{1}} \\ 
(\ell +2\overline{\varepsilon }-\overline{\rho })\overline{\phi _{2}}+(n+%
%TCIMACRO{\U{3bc} }%
%BeginExpansion
\mu
%EndExpansion
-2\gamma )\phi _{0}=(m+2\overline{\pi })\overline{\phi _{1}}+(m-2\tau )\phi
_{1} \\ 
\end{array}
\label{q6}
\end{equation}%
After the substitution of the electron tetrad, I find%
\begin{equation}
\begin{array}{l}
\phi _{0}^{\prime }=\frac{\Delta }{2\overline{\eta }}\phi _{0}\quad ,\quad
\phi _{1}^{\prime }=\frac{1}{\sqrt{2}}\phi _{1}\quad ,\quad \phi
_{2}^{\prime }=\overline{\eta }\phi _{2} \\ 
\overline{m}^{\prime \mu }\partial _{\mu }[\sin \theta \phi _{0}^{\prime
}]-m^{\prime \mu }\partial _{\mu }[\sin \theta \overline{\phi _{0}^{\prime }}%
]=\Delta \sin \theta \lbrack (\ell ^{\prime \mu }\partial _{\mu }+\frac{2}{%
\overline{\eta }})\phi _{1}^{\prime }-(\ell ^{\prime \mu }\partial _{\mu }+%
\frac{2}{\eta })\overline{\phi _{1}^{\prime }}] \\ 
m^{\prime \mu }\partial _{\mu }[\sin \theta \phi _{2}^{\prime }]-\overline{m}%
^{\prime \mu }\partial _{\mu }[\sin \theta \overline{\phi _{2}^{\prime }}%
]=\Delta \sin \theta \lbrack (n^{\prime \mu }\partial _{\mu }-\frac{2}{%
\overline{\eta }})\phi _{1}^{\prime }-(n^{\prime \mu }\partial _{\mu }-\frac{%
2}{\eta })\overline{\phi _{1}^{\prime }}] \\ 
n^{\prime \mu }\partial _{\mu }[\sin \theta \phi _{0}^{\prime }]+\ell
^{\prime \mu }\partial _{\mu }[\sin \theta \overline{\phi _{2}^{\prime }}%
]=\sin \theta \lbrack (m^{\prime \mu }\partial _{\mu }+\frac{2ia\sin \theta 
}{\overline{\eta }})\phi _{1}^{\prime }+(m^{\prime \mu }\partial _{\mu }-%
\frac{2ia\sin \theta }{\eta })\overline{\phi _{1}^{\prime }}]%
\end{array}
\label{q7}
\end{equation}

I will now solve all the equations by separating the real and imaginary
parts. We precisely have for the FE (\ref{q2})%
\begin{equation}
\begin{array}{l}
\phi _{0}^{\prime }=\frac{\Delta }{2\overline{\eta }}\phi _{0}\quad ,\quad
\phi _{1}^{\prime }=\frac{1}{\sqrt{2}}\phi _{1}\quad ,\quad \phi
_{2}^{\prime }=\frac{\rho ^{2}}{\eta }\phi _{2} \\ 
\phi _{0}^{\prime 1}+i\phi _{0}^{\prime 2}=\phi _{0}^{\prime }\sin \theta
\quad ,\quad \phi _{1}^{\prime 1}+i\phi _{1}^{\prime 2}=\phi _{1}^{\prime
}\quad ,\quad \phi _{2}^{\prime 1}+i\phi _{2}^{\prime 2}=\phi _{2}^{\prime
}\sin \theta \\ 
\\ 
\frac{\partial }{\partial \theta }\phi _{0}^{\prime 1}+(\frac{1}{\sin \theta 
}\frac{\partial }{\partial \varphi }+a\sin \theta \frac{\partial }{\partial t%
})\phi _{0}^{\prime 2}=0\quad ,\quad \frac{\partial }{\partial \theta }\phi
_{2}^{\prime 1}-(\frac{1}{\sin \theta }\frac{\partial }{\partial \varphi }%
+a\sin \theta \frac{\partial }{\partial t})\phi _{2}^{\prime 2}=0 \\ 
\frac{\partial }{\partial r}(\phi _{2}^{\prime 1}+\phi _{0}^{\prime 1})+(%
\frac{r^{2}+a^{2}}{\Delta }\frac{\partial }{\partial t}+\frac{a}{\Delta }%
\frac{\partial }{\partial \varphi })(\phi _{2}^{\prime 1}-\phi _{0}^{\prime
1})=0\quad ,\quad \frac{\partial }{\partial r}(\phi _{2}^{\prime 2}-\phi
_{0}^{\prime 2})+(\frac{r^{2}+a^{2}}{\Delta }\frac{\partial }{\partial t}+%
\frac{a}{\Delta }\frac{\partial }{\partial \varphi })(\phi _{2}^{\prime
2}+\phi _{0}^{\prime 2})=0 \\ 
\end{array}
\label{q8}
\end{equation}%
and the integrability conditions (\ref{q7})%
\begin{equation}
\begin{array}{l}
\frac{\partial }{\partial \theta }\phi _{0}^{\prime 2}-(\frac{1}{\sin \theta 
}\frac{\partial }{\partial \varphi }+a\sin \theta \frac{\partial }{\partial t%
})\phi _{0}^{\prime 1}=\Delta \sin \theta \lbrack (\ell ^{\prime \mu
}\partial _{\mu }+\frac{2r}{\eta \overline{\eta }})\phi _{1}^{\prime 2}+%
\frac{2\alpha \cos \theta }{\eta \overline{\eta }}\phi _{1}^{\prime 1}] \\ 
\frac{\partial }{\partial \theta }\phi _{2}^{\prime 2}+(\frac{1}{\sin \theta 
}\frac{\partial }{\partial \varphi }+a\sin \theta \frac{\partial }{\partial t%
})\phi _{2}^{\prime 1}=\Delta \sin \theta \lbrack (n^{\prime \mu }\partial
_{\mu }-\frac{2r}{\eta \overline{\eta }})\phi _{1}^{\prime 2}-\frac{2\alpha
\cos \theta }{\eta \overline{\eta }}\phi _{1}^{\prime 1}] \\ 
\frac{\partial }{\partial r}(\phi _{2}^{\prime 1}-\phi _{0}^{\prime 1})+(%
\frac{r^{2}+a^{2}}{\Delta }\frac{\partial }{\partial t}+\frac{a}{\Delta }%
\frac{\partial }{\partial \varphi })(\phi _{2}^{\prime 1}+\phi _{0}^{\prime
1})=2\sin \theta \lbrack \partial _{\theta }\phi _{1}^{\prime 1}-\frac{%
2\alpha r\sin \theta }{\eta \overline{\eta }}\phi _{1}^{\prime 2}-\frac{%
2\alpha r\sin \theta \cos \theta }{\eta \overline{\eta }}\phi _{1}^{\prime
1}] \\ 
-\frac{\partial }{\partial r}(\phi _{2}^{\prime 2}+\phi _{0}^{\prime 2})-(%
\frac{r^{2}+a^{2}}{\Delta }\frac{\partial }{\partial t}+\frac{a}{\Delta }%
\frac{\partial }{\partial \varphi })(\phi _{2}^{\prime 2}-\phi _{0}^{\prime
2})=2\sin \theta \lbrack (\frac{1}{\sin \theta }\frac{\partial }{\partial
\varphi }+a\sin \theta \frac{\partial }{\partial t})\phi _{1}^{\prime 1}] \\ 
\end{array}
\label{q9}
\end{equation}%
After summing and subtracting the first two rows, they become%
\begin{equation}
\begin{array}{l}
\frac{\partial }{\partial \theta }(\phi _{2}^{\prime 2}+\phi _{0}^{\prime
2})+(\frac{1}{\sin \theta }\frac{\partial }{\partial \varphi }+a\sin \theta 
\frac{\partial }{\partial t})(\phi _{2}^{\prime 1}-\phi _{0}^{\prime
1})=2\Delta \sin \theta \lbrack (\frac{r^{2}+a^{2}}{\Delta }\frac{\partial }{%
\partial t}+\frac{a}{\Delta }\frac{\partial }{\partial \varphi })\phi
_{1}^{\prime 2}] \\ 
\frac{\partial }{\partial \theta }(\phi _{2}^{\prime 2}-\phi _{0}^{\prime
2})+(\frac{1}{\sin \theta }\frac{\partial }{\partial \varphi }+a\sin \theta 
\frac{\partial }{\partial t})(\phi _{2}^{\prime 1}+\phi _{0}^{\prime
1})=-2\Delta \sin \theta \lbrack (\partial _{\theta }+\frac{2r}{\eta 
\overline{\eta }})\phi _{1}^{\prime 2}+\frac{2\alpha \cos \theta }{\eta 
\overline{\eta }}\phi _{1}^{\prime 1}] \\ 
\frac{\partial }{\partial r}(\phi _{2}^{\prime 1}-\phi _{0}^{\prime 1})+(%
\frac{r^{2}+a^{2}}{\Delta }\frac{\partial }{\partial t}+\frac{a}{\Delta }%
\frac{\partial }{\partial \varphi })(\phi _{2}^{\prime 1}+\phi _{0}^{\prime
1})=2\sin \theta \lbrack \partial _{\theta }\phi _{1}^{\prime 1}-\frac{%
2\alpha r\sin \theta }{\eta \overline{\eta }}\phi _{1}^{\prime 2}-\frac{%
2\alpha ^{2}\sin \theta \cos \theta }{\eta \overline{\alpha }}\phi
_{1}^{\prime 1}] \\ 
-\frac{\partial }{\partial r}(\phi _{2}^{\prime 2}+\phi _{0}^{\prime 2})-(%
\frac{r^{2}+a^{2}}{\Delta }\frac{\partial }{\partial t}+\frac{a}{\Delta }%
\frac{\partial }{\partial \varphi })(\phi _{2}^{\prime 2}-\phi _{0}^{\prime
2})=2\sin \theta \lbrack (\frac{1}{\sin \theta }\frac{\partial }{\partial
\varphi }+a\sin \theta \frac{\partial }{\partial t})\phi _{1}^{\prime 1}] \\ 
\end{array}
\label{q10}
\end{equation}

Considering the Fourier transform of $\phi _{\ast }^{\prime \ast }=\underset{%
m\omega }{\dsum }\widetilde{\phi _{\ast }^{\ast }}(\omega ,m;r,\theta
)e^{i\omega t}e^{im\varphi }$ the FE take the form%
\begin{equation}
\begin{array}{l}
\frac{\partial }{\partial \theta }\widetilde{\phi _{0}^{\prime 1}}+iQ%
\widetilde{\phi _{0}^{\prime 2}}=0\quad ,\quad \frac{\partial }{\partial
\theta }\widetilde{\phi _{2}^{\prime 1}}-iQ\widetilde{\phi _{2}^{\prime 2}}%
=0\quad ,\quad Q=\omega a\sin \theta +\frac{m}{\sin \theta } \\ 
\\ 
\frac{\partial }{\partial r}(\widetilde{\phi _{2}^{\prime 1}}+\widetilde{%
\phi _{0}^{\prime 1}})+i\frac{K}{\Delta }(\widetilde{\phi _{2}^{\prime 1}}-%
\widetilde{\phi _{0}^{\prime 1}})=0\quad ,\quad \frac{\partial }{\partial r}(%
\widetilde{\phi _{2}^{\prime 2}}-\widetilde{\phi _{0}^{\prime 2}})+i\frac{K}{%
\Delta }(\widetilde{\phi _{2}^{\prime 2}}+\widetilde{\phi _{0}^{\prime 2}})=0
\\ 
K=\omega (r^{2}+a^{2})+am%
\end{array}
\label{q11}
\end{equation}%
and the integrability conditions become%
\begin{equation}
\begin{array}{l}
\frac{\partial }{\partial \theta }(\widetilde{\phi _{2}^{\prime 2}}+%
\widetilde{\phi _{0}^{\prime 2}})+iQ(\widetilde{\phi _{2}^{\prime 1}}-%
\widetilde{\phi _{0}^{\prime 1}})=2iK\sin \theta \widetilde{\phi
_{1}^{\prime 2}} \\ 
\frac{\partial }{\partial \theta }(\widetilde{\phi _{2}^{\prime 2}}-%
\widetilde{\phi _{0}^{\prime 2}})+iQ(\widetilde{\phi _{2}^{\prime 1}}+%
\widetilde{\phi _{0}^{\prime 1}})=-2\Delta \sin \theta \lbrack (\partial
_{r}+\frac{2r}{\eta \overline{\eta }})\widetilde{\phi _{1}^{\prime 2}}+\frac{%
2\alpha \cos \theta }{\eta \overline{\eta }}\widetilde{\phi _{1}^{\prime 1}}]
\\ 
\frac{\partial }{\partial r}(\widetilde{\phi _{2}^{\prime 1}}-\widetilde{%
\phi _{0}^{\prime 1}})+i\frac{K}{\Delta }(\widetilde{\phi _{2}^{\prime 1}}+%
\widetilde{\phi _{0}^{\prime 1}})=2\sin \theta \lbrack \partial _{\theta }%
\widetilde{\phi _{1}^{\prime 1}}-\frac{2\alpha r\sin \theta }{\eta \overline{%
\eta }}\widetilde{\phi _{1}^{\prime 2}}-\frac{2\alpha ^{2}\sin \theta \cos
\theta }{\eta \overline{\eta }}\widetilde{\phi _{1}^{\prime 1}}] \\ 
-\frac{\partial }{\partial r}(\widetilde{\phi _{2}^{\prime 2}}+\widetilde{%
\phi _{0}^{\prime 2}})-i\frac{K}{\Delta }(\widetilde{\phi _{2}^{\prime 2}}-%
\widetilde{\phi _{0}^{\prime 2}})=2iQ\sin \theta \widetilde{\phi
_{1}^{\prime 1}} \\ 
\end{array}
\label{q12}
\end{equation}%
Making the substitutions%
\begin{equation}
\begin{array}{l}
\widetilde{\phi _{+}^{\prime 1}}=\widetilde{\phi _{2}^{\prime 1}}+\widetilde{%
\phi _{0}^{\prime 1}}\quad ,\quad \widetilde{\phi _{-}^{\prime 1}}=%
\widetilde{\phi _{2}^{\prime 1}}-\widetilde{\phi _{0}^{\prime 1}} \\ 
\widetilde{\phi _{+}^{\prime 2}}=\widetilde{\phi _{2}^{\prime 2}}+\widetilde{%
\phi _{0}^{\prime 2}}\quad ,\quad \widetilde{\phi _{-}^{\prime 2}}=%
\widetilde{\phi _{2}^{\prime 2}}-\widetilde{\phi _{0}^{\prime 2}}%
\end{array}
\label{q13}
\end{equation}%
the FE take the simple form%
\begin{equation}
\begin{array}{l}
\frac{\partial }{\partial \theta }\widetilde{\phi _{+}^{\prime 1}}-iQ%
\widetilde{\phi _{-}^{\prime 2}}=0\quad ,\quad \frac{\partial }{\partial
\theta }\widetilde{\phi _{-}^{\prime 1}}-iQ\widetilde{\phi _{+}^{\prime 2}}=0
\\ 
\frac{\partial }{\partial r}\widetilde{\phi _{+}^{\prime 1}}+i\frac{K}{%
\Delta }\widetilde{\phi _{-}^{\prime 1}}=0\quad ,\quad \frac{\partial }{%
\partial r}\widetilde{\phi _{-}^{\prime 2}}+i\frac{K}{\Delta }\widetilde{%
\phi _{+}^{\prime 2}}=0 \\ 
\\ 
K=\omega (r^{2}+a^{2})+am\quad ,\quad Q=\omega a\sin \theta +\frac{m}{\sin
\theta }%
\end{array}
\label{q14}
\end{equation}%
Notice that only the three from the four relations are independent. The
general solution of these equations is%
\begin{equation}
\begin{array}{l}
\widetilde{\phi _{+}^{\prime 1}}=\widetilde{f}(\omega ,m;r,\theta )\quad
,\quad \widetilde{\phi _{-}^{\prime 1}}=i\frac{\Delta }{K}\frac{\partial 
\widetilde{f}}{\partial r}\quad ,\quad \widetilde{\phi _{-}^{\prime 2}}=-i%
\frac{1}{Q}\frac{\partial \widetilde{f}}{\partial \theta }\quad ,\quad 
\widetilde{\phi _{+}^{\prime 2}}=(\frac{1}{Q}\frac{\partial }{\partial
\theta })(\frac{\Delta }{K}\frac{\partial }{\partial r})\widetilde{f} \\ 
\\ 
\overline{\widetilde{f}(\omega ,m;r,\theta )}=\widetilde{f}(-\omega
,-m;r,\theta )\quad ,\quad f(t,r,\theta ,\varphi )\ is\ \func{real}%
\end{array}
\label{q15}
\end{equation}%
Note that the general solution depends on the general function $\widetilde{f}%
(m,\omega ;r,\theta )$, which will be fixed by the integrability conditions (%
\ref{q12}). Replacing these relations to the integrability conditions, I find%
\begin{equation}
\begin{array}{l}
\widetilde{\phi _{1}^{\prime 1}}=\frac{i}{2Q^{2}\sin \theta }\partial
_{\theta }(\partial _{r}\frac{\Delta }{K}\partial _{r}+\frac{K}{\Delta })%
\widetilde{f}\quad ,\quad \widetilde{\phi _{1}^{\prime 2}}=\frac{-i\Delta }{%
2K^{2}\sin \theta }\partial _{r}(\partial _{\theta }\frac{1}{Q}\partial
_{\theta }-Q)\widetilde{f} \\ 
\\ 
(\frac{\partial }{\partial \theta }\frac{1}{Q}\frac{\partial }{\partial
\theta }-Q)\widetilde{f}=\frac{\Delta \sin \theta }{\eta \overline{\eta }}[-%
\frac{\partial }{\partial r}\frac{\Delta \eta \overline{\eta }}{K^{2}\sin
\theta }\frac{\partial }{\partial r}(\partial _{\theta }\frac{1}{Q}\partial
_{\theta }-Q)\widetilde{f}+\frac{2a\cos \theta }{Q^{2}\sin \theta }\partial
_{\theta }(\partial _{r}\frac{\Delta }{K}\partial _{r}+\frac{K}{\Delta })%
\widetilde{f}] \\ 
(\frac{\partial }{\partial r}\frac{\Delta }{K}\frac{\partial }{\partial r}+%
\frac{K}{\Delta })\widetilde{f}=\frac{\sin \theta }{\eta \overline{\eta }}%
[\partial _{\theta }\frac{\eta \overline{\eta }}{Q^{2}\sin \theta }\partial
_{\theta }(\partial _{r}\frac{\Delta }{K}\partial _{r}+\frac{K}{\Delta })%
\widetilde{f}+\frac{2ar\Delta }{K^{2}}\partial _{r}(\partial _{\theta }\frac{%
1}{Q}\partial _{\theta }-Q)\widetilde{f}%
\end{array}
\label{q16}
\end{equation}%
Notice that\ the last two PDEs are equivalent to the following PDE%
\begin{equation}
\begin{array}{l}
\frac{\eta \overline{\eta }}{\sin \theta }(\partial _{\theta }\frac{1}{Q}%
\partial _{\theta }-Q)(\partial _{r}\frac{\Delta }{K}\partial _{r}+\frac{K}{%
\Delta })\widetilde{f}-\frac{2aK\cos \theta }{Q^{2}\sin \theta }\partial
_{\theta }(\partial _{r}\frac{\Delta }{K}\partial _{r}+\frac{K}{\Delta })%
\widetilde{f}+\frac{2arQ\Delta }{K^{2}}\partial _{r}(\partial _{\theta }%
\frac{1}{Q}\partial _{\theta }-Q)\widetilde{f}=0 \\ 
\end{array}
\label{q17}
\end{equation}%
Using the relation $K-aQ\sin \theta =\eta \overline{\eta }\omega $, it takes
the form%
\begin{equation}
\begin{array}{l}
\frac{K-aQ\sin \theta }{\sin \theta }(\partial _{\theta }\frac{1}{Q}\partial
_{\theta }-Q)(\partial _{r}\frac{\Delta }{K}\partial _{r}+\frac{K}{\Delta })%
\widetilde{f}-\frac{2a\omega K\cos \theta }{Q^{2}\sin \theta }\partial
_{\theta }(\partial _{r}\frac{\Delta }{K}\partial _{r}+\frac{K}{\Delta })%
\widetilde{f}+\frac{2arQ\Delta \omega }{K^{2}}\partial _{r}(\partial
_{\theta }\frac{1}{Q}\partial _{\theta }-Q)\widetilde{f}=0 \\ 
\end{array}
\label{q18}
\end{equation}

This is the final unique PDE that the unknown function $\widetilde{f}$ of (%
\ref{q15}) satisfies. The solutions of this PDE implies solutions for all
the linearized NP components of the gluon field. It is apparent that the
looking for solutions of the form $\widetilde{f}=R(r)S(\theta )$, the
ordinary method of separation of variables does not apply. But notice that
the following two ratios%
\begin{equation}
\begin{array}{l}
\frac{(Q\frac{d}{d\theta }\frac{1}{Q}\frac{d}{d\theta }-Q^{2}-\frac{2a\omega
\cos \theta }{Q}\frac{d}{d\theta })S}{Q\sin \theta (Q\frac{d}{d\theta }\frac{%
1}{Q}\frac{d}{d\theta }-Q^{2})S}=\frac{a(\frac{K}{\Delta }\frac{d}{dr}\frac{%
\Delta }{K}\frac{d}{dr}+\frac{K^{2}}{\Delta ^{2}}-\frac{2ar\omega }{K}\frac{d%
}{dr})R}{K(\frac{K}{\Delta }\frac{d}{dr}\frac{\Delta }{K}\frac{d}{dr}+\frac{%
K^{2}}{\Delta ^{2}})R}=\lambda%
\end{array}
\label{q19}
\end{equation}%
depend on different variables. The first depends on $\theta $ and the second
on $r$. Then as usual, the two ratios must be equal to a constant $\lambda $%
. The derived ODEs are%
\begin{equation}
\begin{array}{l}
(Q\frac{d}{d\theta }\frac{1}{Q}\frac{d}{d\theta }-Q^{2}-\frac{2a\omega \cos
\theta }{Q(1-\lambda Q\sin \theta )}\frac{d}{d\theta })S=0 \\ 
(\frac{K}{\Delta }\frac{d}{dr}\frac{\Delta }{K}\frac{d}{dr}+\frac{K^{2}}{%
\Delta ^{2}}-\frac{2ar\omega }{K(a-\lambda K)}\frac{d}{dr})R=0 \\ 
\\ 
\Delta =r^{2}+a^{2}-2Mr+q^{2}\quad ,\quad K=\omega (r^{2}+a^{2})+am\quad
,\quad Q=\omega a\sin \theta +\frac{m}{\sin \theta }%
\end{array}
\label{q20}
\end{equation}

We see that the NP formalism with the form (\ref{s10}) of the (null) tetrad,
permits the application of the separation of variables, but the implied ODEs
are completely different from those\cite{CHAND} found by Teukolsky for the
electromagnetic field in Kerr spacetime background.

\subsection{Study of the angular differential equation}

In order to compare the present angular ODE with the corresponding $s=0$ ODE
of the Teukolsky master equations, I give it the form%
\begin{equation}
\begin{array}{l}
\frac{1}{\sin \theta }\frac{d}{d\theta }\sin \theta \frac{d}{d\theta }S-%
\frac{2a\omega \cos \theta (2-\lambda Q\sin \theta )}{Q(1-\lambda Q\sin
\theta )}\frac{d}{d\theta }S-Q^{2}S=0 \\ 
\\ 
Q=\omega a\sin \theta +\frac{m}{\sin \theta }%
\end{array}
\label{q21}
\end{equation}%
Notice the difference with the angular $s=0$ Teukolsky master equation%
\begin{equation}
\begin{array}{l}
\frac{1}{\sin \theta }\frac{d}{d\theta }\sin \theta \frac{d}{d\theta }%
S+(E+a^{2}\omega ^{2}\cos ^{2}\theta -\frac{m^{2}}{\sin ^{2}\theta })S=0 \\ 
\end{array}
\label{q22}
\end{equation}

The essential difference is that (\ref{q22}) is a smooth deformation (with
parameter $a$) to the ordinary spherical harmonics. For $a=0$ the ODE (\ref%
{q21}) takes the form 
\begin{equation}
\begin{array}{l}
\frac{1}{\sin \theta }\frac{d}{d\theta }\sin \theta \frac{d}{d\theta }S-%
\frac{m^{2}}{\sin ^{2}\theta }S=0 \\ 
\end{array}
\label{q23}
\end{equation}%
which is apparently singular for $m\neq 0$. The above equation (\ref{q23})
has the form of the differential equation of the spherical harmonics with $%
l=0$. This implies that only the constant spherical harmonic $Y_{00}$ can be
deformed to a regular solution of the ODE (\ref{q21}). This imposes a very
strong restriction to this ODE, which we have to solve only for $m=0$. In
this case and for $a\omega \neq 0$, the initial ODE (\ref{q20}) takes the
form%
\begin{equation}
\begin{array}{l}
\lbrack \frac{1}{\sin \theta }\frac{d}{d\theta }\frac{1}{\sin \theta }\frac{d%
}{d\theta }-\frac{2\cos \theta }{\sin ^{3}\theta (1-\lambda a\omega \sin
^{2}\theta )}\frac{d}{d\theta }-(a\omega )^{2}]S=0 \\ 
\end{array}
\label{q24}
\end{equation}%
Using the variable $c=-\cos \theta $, it becomes%
\begin{equation}
\begin{array}{l}
\lbrack \frac{d^{2}}{dc^{2}}+(\frac{2c}{1-c^{2}}+\frac{2\lambda a\omega c}{%
1-\lambda a\omega +\lambda a\omega c^{2}})\frac{d}{dc}-(a\omega )^{2}]S=0 \\ 
c=-\cos \theta \quad ,\quad \lbrack 0\ ,\ \pi ]\rightarrow \lbrack -1\ ,\ +1]%
\end{array}
\label{q25}
\end{equation}%
Hiding the first derivative into $S$, through the transformation%
\begin{equation}
\begin{array}{l}
S=\Phi \exp \dint \kappa dc \\ 
\end{array}
\label{q26}
\end{equation}%
I find%
\begin{equation}
\begin{array}{l}
\lbrack \frac{d^{2}}{dc^{2}}-\frac{d\kappa }{dc}-\kappa ^{2}-(a\omega
)^{2}]\Phi =0 \\ 
\\ 
\kappa =\frac{-c}{1-c^{2}}+\frac{-\lambda a\omega c}{1-\lambda a\omega
+\lambda a\omega c^{2}} \\ 
S=\Phi \exp \dint \kappa dc=\frac{1-c^{2}}{2(1-\lambda a\omega +\lambda
a\omega c^{2})}\Phi%
\end{array}
\label{q27}
\end{equation}%
That is%
\begin{equation}
\begin{array}{l}
\lbrack \frac{d^{2}}{dc^{2}}-U]\Phi =0 \\ 
U(c)=\frac{1+2c^{2}}{(1-c^{2})^{2}}+\frac{\lambda a\omega (1-\lambda a\omega
)}{(1-\lambda a\omega +\lambda a\omega c^{2})^{2}}+\frac{2\lambda a\omega
c^{2}}{(1-c^{2})(1-\lambda a\omega +\lambda a\omega c^{2})}+(a\omega )^{2}
\\ 
\end{array}
\label{q28}
\end{equation}

The function $U(c)$ becomes infinite at the boundaries $c=\pm 1$. It also
diverges to $-\infty $ for%
\begin{equation}
\begin{array}{l}
0<c^{2}=\frac{\lambda \omega a-1}{\lambda \omega a}<1 \\ 
\end{array}
\label{q29}
\end{equation}%
which is possible if $\lambda \omega a>1$. Apparently the search of
solutions of $\omega $ and $\lambda $ have to be done in combination with
the possible solutions of the radial ODE. Taking into account the complexity
of the function $U(c)$, the WKB approximation is indicated to be used, in
order to find numerical results.

\subsection{Study of the radial differential equation}

Analyzing the angular ODE, I found the condition $m=0$, which we have to
impose to the radial ODE (\ref{q20}). I finally find 
\begin{equation}
\begin{array}{l}
(\frac{d^{2}}{dr^{\prime 2}}+\omega ^{2}-2\sigma \frac{d}{dr^{\prime }}%
)R=0\quad ,\quad \sigma =\frac{ar\Delta }{(r^{2}+a^{2})^{2}(a-\lambda \omega
(r^{2}+a^{2}))} \\ 
\\ 
r^{\prime }=r-\frac{2M^{2}-q^{2}}{d}\arctan \frac{d}{r-M}+M\ln \Delta
+C\quad ,\quad d=\sqrt{a^{2}-M^{2}+q^{2}}%
\end{array}
\label{q30}
\end{equation}%
where I have also made a change of variable such that $\frac{dr^{\prime }}{dr%
}=\frac{r^{2}+a^{2}}{\Delta }$. After making the transformation%
\begin{equation}
\begin{array}{l}
R(r^{\prime })=\Psi (r^{\prime })\exp \dint \sigma (r(r^{\prime
}))dr^{\prime }\quad ,\quad \sigma (r)=\frac{ar\Delta }{(r^{2}+a^{2})^{2}(a-%
\lambda \omega (r^{2}+a^{2}))} \\ 
\end{array}
\label{q31}
\end{equation}%
the 1st derivative term is removed and the radial ODE takes the form%
\begin{equation}
\begin{array}{l}
\lbrack \frac{d^{2}}{dr^{\prime 2}}-V(r(r^{\prime }))]\Psi (r^{\prime })=0
\\ 
V(r)=\sigma ^{2}-\frac{\Delta }{r^{2}+a^{2}}\frac{d\sigma }{dr}-\omega
^{2}\quad ,\quad \sigma (r)=\frac{ar\Delta }{(r^{2}+a^{2})^{2}(a-\lambda
\omega (r^{2}+a^{2}))} \\ 
\end{array}
\label{q32}
\end{equation}

For asymptotic values of $r$ we have%
\begin{equation}
\begin{array}{l}
\sigma \simeq -\frac{a}{\lambda \omega r^{3}}\quad ,\quad V(r)\simeq \frac{3a%
}{\lambda \omega r^{4}}-\omega ^{2} \\ 
\Psi (r)\simeq e^{\pm i\omega r} \\ 
\end{array}
\label{q33}
\end{equation}%
This "free" gluon behavior should not confuse us, because it is a solution
of the linear part of the gluonic equation.

The "potentials" $U(c)$ and $V(r)$\ are too complicated to find formal
solutions of the angular and radial ODEs. In order to find the "spectrum" of
(\ref{q28}) and (\ref{q32}) ODEs, i.e. the possible solutions of $\omega $
and $\lambda $,\ we have to make numerical calculations using the ordinary
WKB approximation.

\section{COLORED SOLITONS}

\setcounter{equation}{0}

At the classical level, the four dimensional LCR-structure tetrad defines a
class of symmetric tensors $g_{\mu \nu }$ and antisymmetric self-dual
tensors $J_{\mu \nu }$. We have already made clear (\ref{s12}) that
classical electrodynamics is generated for the case of the static
LCR-manifold because there is a function $f$ such that $d(fJ)=i\ast j$,
where $j$ is a distributional source. Einstein's gravity is generated from
the corresponding asymptotically flat symmetric tensor, which defines the
conserved energy-momentum and angular-momentum in the linearized
approximation. I have also made clear\cite{RAG2018a} that the massive
LCR-soliton (electron) depends on left and right hand chiral parts, which
are defined by the two $CP^{3}$ intersection points of the irreducible
quadratic hypersurface and a line. On the other hand the massless
LCR-soliton (neutrino) of the reducible quadratic hypersurface also depends
on left and right hand chiral parts, but one of them is degenerate. It
coincides with the corresponding part of the (degenerate) vacuum
LCR-structure. In the context of solitonic framework, the description of the
correspondence between leptons and quarks needs a classical solution of the
present gauge field equation in the static LCR-manifold background with a
distributional source, which could be identified with the quark. In the
present subsection I will derive such a gauge field solution from the action
(\ref{i7a}), in complete analogy to electromagnetic field derivation.

Both lagrangians (\ref{i7a}) and (\ref{i7}) explicitly contain a gauge field
compatible with the LCR-structure, but it is (\ref{i7a}) that admits
distributional solitons. Therefore in this section we will focus on this
action, Variation of the action relative to this gauge field implies the
field equations 
\begin{equation}
\begin{array}{l}
\frac{1}{\sqrt{-g}}(D_{\mu })_{ij}(\sqrt{-g}(\Gamma ^{\mu \nu \rho \sigma }-%
\overline{\Gamma ^{\mu \nu \rho \sigma }})F_{j\rho \sigma })=0 \\ 
\\ 
\Gamma ^{\mu \nu \rho \sigma }=\frac{1}{2}[(\ell ^{\mu }m^{\nu }-\ell ^{\nu
}m^{\mu })(n^{\rho }\overline{m}^{\sigma }-n^{\sigma }\overline{m}^{\rho
})+(n^{\mu }\overline{m}^{\nu }-n^{\nu }\overline{m}^{\mu })(\ell ^{\rho
}m^{\sigma }-\ell ^{\sigma }m^{\rho })] \\ 
\end{array}
\label{f1}
\end{equation}%
where $(D_{\mu })_{ij}=\delta _{ij}\partial _{\mu }-\gamma f_{ikj}A_{k\mu }$
is the gauge symmetry covariant derivative and $\gamma $ the coupling
constant. The derivation of quantum electrodynamics as an affective field
theory was triggered by the existence of a source in the closed self-dual
antisymmetric tensor of the massive static soliton. But the above (\ref{f1})
field equation is exact. We cannot replace (ad hoc) the zero of the second
part of the equation with a source, because the symmetries of the action
will be destroyed.

I first notice that (\ref{f1}) may be written in the form 
\begin{equation}
\begin{array}{l}
\frac{1}{\sqrt{-g}}(D_{\mu })_{ij}\{\sqrt{-g}[(\ell ^{\mu }m^{\nu }-\ell
^{\nu }m^{\mu })(n^{\rho }\overline{m}^{\sigma }F_{j\rho \sigma })+ \\ 
\qquad +(n^{\mu }\overline{m}^{\nu }-n^{\nu }\overline{m}^{\mu })(\ell
^{\rho }m^{\sigma }F_{j\rho \sigma })]\}=-k_{i}^{\nu } \\ 
\end{array}
\label{f2}
\end{equation}%
where $k_{i}^{\nu }(x)$ is a real vector field and I assume that the
LCR-manifold is fixed to be the static one. This is possible, because the
LCR-structure equations completely decouple from the hadronic sector. The
LCR-structure is first fixed and after we proceed to the solution of the
field equations which involve the gluonic field. This property is
essentially behind the physical observation of the lepton-quark
correspondence!

A classical solution of the gauge field with a compact source will be
interpreted as a colored soliton (the quark) with its gluon potential. If we
apply with the gauge covariant derivative $(D_{\nu })_{ij}$ and use the
commutation relation 
\begin{equation}
\begin{array}{l}
\lbrack (D_{\mu }),(D_{\nu })]_{ik}=-\gamma f_{ijk}F_{j\mu \nu } \\ 
\end{array}
\label{f3}
\end{equation}%
we find that the current must be covariantly conserved $(D_{\nu
})_{ij}k_{j}^{\nu }=0$ for a classical solution to exist.

The reduction of the original real field equation (\ref{f1}) to the two
equivalent complex (\ref{f2}) PDE with real source permit us to look for
solutions in the ambient complex manifold and after reduce them down to the
LCR-manifold via the structure functions ($z^{\alpha }(x),z^{\widetilde{%
\beta }}(x)$). But the ambient complex manifold is the space of double
points on an algebraic hypersurface of $CP^{3}$ (i.e. its intersections with
the lines). That is we look for solutions of the equations 
\begin{equation}
\begin{array}{l}
dG^{(+)}-\gamma \lbrack A,G^{(+)}]=i\ast k \\ 
\\ 
G_{j}^{(+)}\equiv (n^{\rho }\overline{m}^{\sigma }F_{j\rho \sigma })\ell
\wedge m+(\ell ^{\rho }m^{\sigma }F_{j\rho \sigma })n\wedge \overline{m}%
\end{array}
\label{f4}
\end{equation}%
where the generally complex current will now be related to the singular
points of the algebraic hypersurface.

In order to clarify the procedure, I first consider the case of an abelian
gauge group ($\gamma =0$) and the simplified PDEs 
\begin{equation}
\begin{array}{l}
d[(n^{\rho }\widetilde{m}^{\sigma }F_{j\rho \sigma })\ell \wedge m+(\ell
^{\rho }m^{\sigma }F_{j\rho \sigma })n\wedge \widetilde{m}]=i\ast k \\ 
\end{array}
\label{f5}
\end{equation}%
The following general relations between the structure coordinates and the
complexified tetrad%
\begin{equation}
\begin{array}{l}
dz^{\alpha }=f_{\alpha }\ \ell _{\mu }dx^{\mu }+h_{\alpha }\ m_{\mu }dx^{\mu
}\;\;\;\;,\;\;\;dz^{\widetilde{\alpha }}=f_{\widetilde{\alpha }}\ n_{\mu
}dx^{\mu }+h_{\widetilde{\alpha }}\ \widetilde{m}_{\mu }dx^{\mu } \\ 
\\ 
\ell =\ell _{\alpha }dz^{\alpha }\;\;\;\;,\;\;\;m=m_{\alpha }dz^{\alpha } \\ 
n=n_{\widetilde{\alpha }}dz^{\widetilde{\alpha }}\;\;\;\;,\;\;\;\widetilde{m}%
=\widetilde{m}_{\widetilde{\alpha }}dz^{\widetilde{\alpha }} \\ 
\end{array}
\label{f6}
\end{equation}%
implies the following closed 2-forms%
\begin{equation}
\begin{array}{l}
dz^{0}\wedge dz^{1}=(f_{0}h_{1}-f_{1}h_{0})\ell \wedge m \\ 
\\ 
dz^{\widetilde{0}}\wedge dz^{\widetilde{1}}=(f_{\widetilde{0}}h_{\widetilde{1%
}}-f_{\widetilde{1}}h_{\widetilde{0}})n\wedge \widetilde{m} \\ 
\end{array}
\label{f7}
\end{equation}%
That is, the two 2-forms $\ell \wedge m$ and $n\wedge \widetilde{m}$ always
admit functions which make them closed. The static LCR-tetrad (\ref{s9})
admits the structure coordinates%
\begin{equation}
\begin{array}{l}
z^{0}=t-f_{0}(r)+ia\cos \theta \quad ,\quad z^{1}=e^{i\varphi
}e^{-iaf_{1}(r)}\tan \frac{\theta }{2} \\ 
z^{\widetilde{0}}=t+f_{0}(r)-ia\cos \theta \quad ,\quad z^{\widetilde{1}%
}=e^{-i\varphi }e^{-iaf_{1}(r)}\tan \frac{\theta }{2} \\ 
\\ 
f_{0}(r)=\tint \frac{r^{2}+a^{2}}{\Delta }dr\quad ,\quad f_{1}(r)=\tint 
\frac{1}{\Delta }dr%
\end{array}
\label{f7a}
\end{equation}%
and the corresponding tetrad components%
\begin{equation}
\begin{array}{l}
\ell _{\mu }dx^{\mu }=dz^{0}+\frac{2ia\sin \theta \cos ^{2}\frac{\theta }{2}%
}{e^{i\varphi }e^{-iaf_{1}(r)}}dz^{1} \\ 
m_{\mu }dx^{\mu }=\frac{ia\sin \theta }{\eta \sqrt{2}}dz^{0}-\frac{\sqrt{2}%
(r^{2}+a^{2})\cos ^{2}\frac{\theta }{2}}{\eta e^{i\varphi }e^{-iaf_{1}(r)}}%
dz^{1} \\ 
\\ 
n_{\mu }dx^{\mu }=dz^{\widetilde{0}}-\frac{ia\sin \theta \cos ^{2}\frac{%
\theta }{2}}{e^{-i\varphi }e^{-iaf_{1}(r)}\eta \overline{\eta }}dz^{%
\widetilde{1}} \\ 
\overline{m}_{\mu }dx^{\mu }=-\frac{ia\sin \theta }{\overline{\eta }\sqrt{2}}%
dz^{\widetilde{0}}-\frac{\sqrt{2}(r^{2}+a^{2})\cos ^{2}\frac{\theta }{2}}{%
\overline{\eta }e^{-i\varphi }e^{-iaf_{1}(r)}}dz^{\widetilde{1}}%
\end{array}
\label{f7b}
\end{equation}%
Recall that the structure coordinates are the coordinates of two points on
two different sheets of the static form of the regular quadric. The implied
closed forms correspond to the non-vanishing holomorphic form, which is
constant, because the quadric is compact.

The closed 2-forms are

\begin{equation}
\begin{array}{l}
d(\frac{C^{\prime }}{\sin \theta (r-ia\cos \theta )}\ell \wedge m)=0 \\ 
\\ 
d(\frac{C^{\prime \prime }(r-ia\cos \theta )}{(r^{2}-2Mr+a^{2}+q^{2})\sin
\theta }n\wedge \widetilde{m})=0 \\ 
\end{array}
\label{f10}
\end{equation}%
where $C^{\prime }$ and $C^{\prime \prime }$ are arbitrary complex
constants. They have sources because Stoke's theorem

\begin{equation}
\begin{array}{l}
\tint\limits_{t,r=const}\frac{C^{\prime }}{\sin \theta (r-ia\cos \theta )}%
\ell \wedge m=-2\sqrt{2}\pi C^{\prime }a \\ 
\\ 
\tint\limits_{t,r=const}\frac{C^{\prime \prime }(r-ia\cos \theta )}{%
(r^{2}-2Mr+a^{2}+q^{2})\sin \theta }n\wedge \overline{m}=-\sqrt{2}\pi
C^{\prime \prime }a%
\end{array}
\label{f11}
\end{equation}%
Notice that they are proportional to the coefficient $a$ implying that the
rotationally symmetric LCR-structure does not define colored configurations
with sources.

Hence a solution is 
\begin{equation}
\begin{array}{l}
(\ell ^{\rho }m^{\sigma }F_{j\rho \sigma })=\frac{C_{j}^{\prime \prime
}(r-ia\cos \theta )}{(r^{2}-2Mr+a^{2}+q^{2})\sin \theta }\quad ,\quad
(n^{\rho }\overline{m}^{\sigma }F_{j\rho \sigma })=\frac{C_{j}^{\prime }}{%
\sin \theta (r-ia\cos \theta )} \\ 
\end{array}
\label{f12}
\end{equation}%
with "electric" and "magnetic" sources, because the constants are generally
complex. But the current $k^{\nu }(x)$ must be real for the field equations
to be satisfied. This condition finally fix 
\begin{equation}
\begin{array}{l}
2C_{j}^{\prime }+C_{j}^{\prime \prime }=-\frac{\gamma _{j}}{\pi a\sqrt{2}}
\\ 
\end{array}
\label{f13}
\end{equation}%
because of the relations 
\begin{equation}
\begin{array}{l}
\frac{C_{j}}{\sin \theta (r-ia\cos \theta )}\ell \wedge m+\frac{%
C_{j}^{\prime }(r-ia\cos \theta )}{(r^{2}-2Mr+a^{2}+q^{2})\sin \theta }%
n\wedge \overline{m}=G_{j}-i\ast G_{j} \\ 
\\ 
\tint\limits_{t,r=const}[\frac{C_{j}^{\prime }}{\sin \theta (r-ia\cos \theta
)}\ell \wedge m+\frac{C_{j}^{\prime \prime }(r-ia\cos \theta )}{%
(r^{2}-2Mr+a^{2}+q^{2})\sin \theta }n\wedge \overline{m}]\equiv \gamma _{j}
\\ 
\end{array}
\label{f14}
\end{equation}%
where $\gamma _{j}$ are independent constants of the abelean real color
charges.

\subsection{A non-abelian solution}

At this level I consider the complexification of $x$, which is necessary for
the application of the holomorphic Frobenious theorem and the solution of
the LCR-structure integrability conditions. But after this complexification,
the PDE (\ref{f2}) permit us to make a holomorphic transformation to the
structure coordinates ($z^{\alpha }(x),z^{\widetilde{\alpha }}(x)$) implied
by the complexified LCR-tetrad ($\ell ,m;n,\widetilde{m}$) 
\begin{equation}
\begin{array}{l}
\frac{1}{\sqrt{-g}}(\partial _{a}\delta _{ij}-\gamma f_{ikj}A_{ka})\{\sqrt{-g%
}[(\ell ^{a}m^{b}-\ell ^{b}m^{a})(n^{c}\widetilde{m}^{d}F_{jcd})+ \\ 
\qquad +(n^{a}\widetilde{m}^{b}-n^{b}\widetilde{m}^{a})(\ell
^{c}m^{d}F_{jcd})\}=-k_{i}^{b} \\ 
\end{array}
\label{f15}
\end{equation}%
which is considerably simplified, because in the structure coordinates
(which are holomorphic functions of complexified $x$) the following
relations hold

\begin{equation}
\begin{array}{l}
dz^{\alpha }=f_{0}^{\alpha }\ \ell _{\mu }dx^{\mu }+f_{1}^{\alpha }\ m_{\mu
}dx^{\mu }\;\;\;\;,\;\;\;dz^{\widetilde{\alpha }}=f_{\widetilde{0}}^{%
\widetilde{\alpha }}\ n_{\mu }dx^{\mu }+f_{\widetilde{1}}^{\widetilde{\alpha 
}}\ \overline{m}_{\mu }dx^{\mu } \\ 
\\ 
\ell _{\mu }dx^{\mu }=\ell _{\alpha }dz^{\alpha }\;\;\;\;,\;\;\;m_{\mu
}dx^{\mu }=m_{\alpha }dz^{\alpha }\;\;\;\;,\;\;\;n_{\mu }dx^{\mu }=n_{%
\widetilde{\alpha }}dz^{\widetilde{\alpha }}\;\;\;\;,\;\;\;\overline{m}_{\mu
}dx^{\mu }=\overline{m}_{\widetilde{\alpha }}dz^{\widetilde{\alpha }} \\ 
\ell ^{\mu }\partial _{\mu }=\ell ^{\widetilde{\alpha }}\partial _{%
\widetilde{\alpha }}\;\;\;\;,\;\;\;m^{\mu }\partial _{\mu }=m^{\widetilde{%
\alpha }}\partial _{\widetilde{\alpha }}\;\;\;\;,\;\;\;n^{\mu }\partial
_{\mu }=n^{\alpha }\partial _{\alpha }\;\;\;\;,\;\;\;\overline{m}^{\mu
}\partial _{\mu }=\overline{m}^{\alpha }\partial _{\alpha } \\ 
\end{array}
\label{f16}
\end{equation}%
and

\begin{equation}
\begin{array}{l}
ds^{2}=g_{\mu \nu }dx^{\mu }dx^{\nu }=g_{ab}dz^{a}dz^{b}=2g_{\alpha 
\widetilde{\beta }}dz^{\alpha }dz^{\widetilde{\beta }}=2(\ell _{\alpha }n_{%
\widetilde{\beta }}-m_{\alpha }\overline{m}_{\widetilde{\alpha }})dz^{\alpha
}dz^{\widetilde{\beta }} \\ 
\\ 
g_{ab}=%
\begin{pmatrix}
0 & g_{\alpha \widetilde{\beta }} \\ 
g_{\beta \widetilde{\alpha }} & 0%
\end{pmatrix}%
\;\;\;\;,\;\;\;g^{ab}=%
\begin{pmatrix}
0 & g^{\alpha \widetilde{\beta }} \\ 
g^{\beta \widetilde{\alpha }} & 0%
\end{pmatrix}
\\ 
g_{\alpha \widetilde{\beta }}=\ell _{\alpha }n_{\widetilde{\beta }%
}-m_{\alpha }\overline{m}_{\widetilde{\beta }}\;\;\;\;,\;\;\;g^{\alpha 
\widetilde{\beta }}=n^{\alpha }\ell ^{\widetilde{\beta }}-\overline{m}%
^{\alpha }m^{\widetilde{\beta }}\;\;\;,\;\;\;g=\det g_{ab}=\widetilde{g}^{2}
\\ 
(\ell _{0}m_{1}-m_{0}\ell _{1})(n_{\widetilde{0}}\overline{m}_{\widetilde{1}%
}-\overline{m}_{\widetilde{0}}n_{\widetilde{1}})=-\widetilde{g}%
\;\;\;,\;\;\;(\ell ^{\widetilde{0}}m^{\widetilde{1}}-m^{\widetilde{0}}\ell ^{%
\widetilde{1}})(n^{0}\overline{m}^{1}-\overline{m}^{0}n^{1})=-\frac{1}{%
\widetilde{g}}%
\end{array}
\label{f17}
\end{equation}%
Besides, the used orientation yields the following formulas

\begin{equation}
\begin{array}{l}
\sqrt{-g}dx^{0}\wedge dx^{1}\wedge dx^{2}\wedge dx^{3}=-i\ell \wedge m\wedge
n\wedge \widetilde{m}=-i\widehat{g}dz^{0}\wedge dz^{1}\wedge dz^{\widehat{0}%
}\wedge dz^{\widehat{1}} \\ 
\\ 
g_{ab}=%
\begin{pmatrix}
0 & \widehat{g}_{\alpha \widetilde{\beta }} \\ 
\widehat{g}_{\beta \widetilde{\alpha }} & 0%
\end{pmatrix}%
\;\;\;\;,\;\;\;g^{ab}=%
\begin{pmatrix}
0 & \widehat{g}^{\alpha \widetilde{\beta }} \\ 
\widehat{g}^{\beta \widetilde{\alpha }} & 0%
\end{pmatrix}
\\ 
\ \widehat{g}_{\alpha \widetilde{\beta }}=\ell _{\alpha }n_{\widetilde{\beta 
}}-m_{\alpha }\overline{m}_{\widetilde{\beta }}\;\;\;\;,\;\;\;\widehat{g}%
^{\alpha \widetilde{\beta }}=n^{\alpha }\ell ^{\widetilde{\beta }}-\overline{%
m}^{\alpha }m^{\widetilde{\beta }}\;\;\;,\;\;\;\widehat{g}\equiv \det 
\widehat{g}_{\alpha \widetilde{\beta }} \\ 
(\ell _{0}m_{1}-m_{0}\ell _{1})(n_{\widetilde{0}}\overline{m}_{\widetilde{1}%
}-\overline{m}_{\widetilde{0}}n_{\widetilde{1}})=-\widehat{g}%
\;\;\;,\;\;\;(\ell ^{\widetilde{0}}m^{\widetilde{1}}-m^{\widetilde{0}}\ell ^{%
\widetilde{1}})(n^{0}\overline{m}^{1}-\overline{m}^{0}n^{1})=-\frac{1}{%
\widehat{g}}%
\end{array}
\label{f18}
\end{equation}%
Hence after the complexification we have to replace $\sqrt{-g}\rightarrow -i%
\widehat{g}$. Notice that now we deal with a complex metric (pseudo-metric),
and we must not take complex conjugations before returning back to real $x$.
I find that (\ref{f15}) imply the PDEs 
\begin{equation}
\begin{array}{l}
For\ b=0\quad ,\quad \partial _{1}F_{i\widetilde{0}\widetilde{1}}-\gamma
f_{ikj}A_{k1}F_{j\widetilde{0}\widetilde{1}}=(D_{1})_{ij}F_{j\widetilde{0}%
\widetilde{1}}=-\widehat{g}k_{i}^{0} \\ 
For\ b=1\quad ,\quad \partial _{0}F_{i\widetilde{0}\widetilde{1}}-\gamma
f_{ikj}A_{k0}F_{j\widetilde{0}\widetilde{1}}=(D_{0})_{ij}F_{j\widetilde{0}%
\widetilde{1}}=\widehat{g}k_{i}^{1} \\ 
For\ b=\widetilde{0}\quad ,\quad \partial _{\widetilde{1}}F_{i01}-\gamma
f_{ikj}A_{k\widetilde{1}}F_{j01}=(D_{\widetilde{1}})_{ij}F_{j01}=-\widehat{g}%
k_{i}^{\widetilde{0}} \\ 
For\ b=\widetilde{1}\quad ,\quad \partial _{\widetilde{0}}F_{i01}-\gamma
f_{ikj}A_{k\widetilde{0}}F_{j01}=(D_{\widetilde{0}})_{ij}F_{j01}=\widehat{g}%
k_{i}^{\widetilde{1}} \\ 
\end{array}
\label{f19}
\end{equation}
The integrability conditions 
\begin{equation}
\begin{array}{l}
\lbrack (D_{0}),(D_{1})]_{ik}F_{k\widetilde{0}\widetilde{1}}=-\gamma
f_{ijk}F_{j01}F_{k\widetilde{0}\widetilde{1}}=-(D_{\beta })_{ij}(\widehat{g}%
k_{j}^{\beta }) \\ 
\lbrack (D_{\widetilde{0}}),(D_{\widetilde{1}})]_{ik}F_{k01}=-\gamma
f_{ijk}F_{j\widetilde{0}\widetilde{1}}F_{k01}=-(D_{\widetilde{\beta }})_{ij}(%
\widehat{g}k_{j}^{\widetilde{\beta }}) \\ 
\\ 
(D_{\beta })_{ij}(\widehat{g}k_{j}^{\beta })+(D_{\widetilde{\beta }})_{ij}(%
\widehat{g}k_{j}^{\widetilde{\beta }})=(D_{b})_{ij}(\widehat{g}k_{j}^{b})=0
\\ 
\end{array}
\label{f20}
\end{equation}%
yield the expected covariantly conserved current relation.

As expected, the written in LCR-structure coordinates PDEs (\ref{f19}) do
not contain the complexified "metric" $g_{\alpha \widetilde{\beta }}$,\ and
contain only the self-dual components $F_{j01}$\ and $F_{j\widetilde{0}%
\widetilde{1}}$ of the gauge field strength, because the present gauge field
action has been constructed to be metric independent.

It is evident that if 
\begin{equation}
\begin{array}{l}
f_{ijk}F_{j01}F_{k\widetilde{0}\widetilde{1}}=-\frac{1}{\widehat{g}}%
f_{ijk}(n^{\mu }\widetilde{m}^{\nu }F_{j\mu \nu })(\ell ^{\mu }m^{\nu
}F_{k\mu \nu })\neq 0 \\ 
\end{array}
\label{f21}
\end{equation}%
does not vanish, the differential equations (\ref{f19} \ do not accept
solutions outside the compact support of the sources. There is an internal
contradiction. Hence my conclusion is that we may only have solutions if $%
F_{i01}$\ or $F_{j\widetilde{0}\widetilde{1}}$ vanish for non vanishing $%
f_{ijk}$. That is we may have the following solutions\ 
\begin{equation}
\begin{array}{l}
A_{\alpha }=\frac{1}{\gamma }(\partial _{\alpha }U)U^{-1}\quad ,\quad (\ell
^{\mu }m^{\nu }F_{k\mu \nu })=(\ell ^{\widetilde{0}}m^{\widetilde{1}}-m^{%
\widetilde{0}}\ell ^{\widetilde{1}})F_{k\widetilde{0}\widetilde{1}}\neq 0 \\ 
\\ 
A_{\widetilde{\alpha }}=\frac{1}{\gamma }(\partial _{\widetilde{\alpha }%
}U^{\prime })U^{\prime -1}\quad ,\quad (n^{\mu }\widetilde{m}^{\nu }F_{k\mu
\nu })=(n^{0}\overline{m}^{1}-\overline{m}^{0}n^{1})F_{k01}\neq 0 \\ 
\end{array}
\label{f22}
\end{equation}%
where $U$ and $U^{\prime }$ are arbitrary elements of the gauge group in a
prescribed representation.

Going back to the PDEs (\ref{f19}) and using the Lie algebra form of the
gauge potential and field 
\begin{equation}
\begin{array}{l}
F\equiv dA-\gamma A\wedge A\quad ,\quad dF-\gamma A\wedge F+\gamma F\wedge
A\equiv 0 \\ 
\\ 
A=A_{j\mu }\tau _{j}dx^{\mu }\quad ,\quad \lbrack \tau _{j},\tau
_{k}]=f_{jki}\tau _{i}\quad ,\quad tr(\tau _{j}\tau _{k})=-\frac{\delta _{jk}%
}{2}\quad ,\quad (\tau _{j})^{\dag }=-\tau _{j} \\ 
\end{array}
\label{f23}
\end{equation}%
the first case gives 
\begin{equation}
\begin{array}{l}
\partial _{\alpha }F_{\widetilde{0}\widetilde{1}}-\gamma \lbrack A_{\alpha
},F_{\widetilde{0}\widetilde{1}}]=0 \\ 
\\ 
\partial _{\alpha }F_{\widetilde{0}\widetilde{1}}^{\prime }=0\quad ,\quad F_{%
\widetilde{0}\widetilde{1}}=UF_{\widetilde{0}\widetilde{1}}^{\prime }U^{-1}
\\ 
\end{array}
\label{f24}
\end{equation}%
which apparently coincides with the abelian PDE (\ref{f9}) 
\begin{equation}
\begin{array}{l}
d((\ell ^{\mu }m^{\nu }F_{k\mu \nu })n\wedge \overline{m})=0\quad ,\quad
(n^{\mu }\overline{m}^{\nu }F_{k\mu \nu })=0 \\ 
\end{array}
\label{f25}
\end{equation}%
which has already been solved. Hence the solution is 
\begin{equation}
\begin{array}{l}
(\ell ^{\mu }m^{\nu }F_{k\mu \nu })=\frac{C_{j}^{\prime \prime }(r-ia\cos
\theta )}{(r^{2}-2Mr+a^{2}+q^{2})\sin \theta }\quad ,\quad (n^{\mu }%
\overline{m}^{\nu }F_{k\mu \nu })=0 \\ 
\end{array}
\label{f26}
\end{equation}%
with arbitrary constants $C_{j}^{\prime \prime }$, which, combined with the
arbitrary group element $U(x)$, yield $\phi _{j0}$.

Assuming $(\ell ^{\mu }m^{\nu }F_{k\mu \nu })=0$, the second solution is 
\begin{equation}
\begin{array}{l}
(\ell ^{\mu }m^{\nu }F_{k\mu \nu })=0\quad ,\quad (n^{\mu }\overline{m}^{\nu
}F_{k\mu \nu })=\frac{C_{j}^{\prime }}{\sin \theta (r-ia\cos \theta )} \\ 
\end{array}
\label{f27}
\end{equation}%
with arbitrary constants $C_{j}^{\prime }$.

In zero gravity, the non-vanishing solutions of the field strength (with
their potentials) are found to be

\begin{equation}
\begin{array}{l}
F_{j}^{\prime }=\frac{-\gamma _{j}^{\prime }}{\pi a\sqrt{2}}[\frac{a}{%
r^{2}+a^{2}}dt\wedge dr-d(t-r)\wedge d\varphi ]= \\ 
\qquad =d[\frac{-\gamma _{j}}{\pi a\sqrt{2}}(t-r)(\frac{a}{r^{2}+a^{2}}%
dr-d\varphi )] \\ 
\\ 
F_{j}^{\prime \prime }=\frac{-\gamma _{j}^{\prime \prime }\sqrt{2}}{\pi a}[%
\frac{a}{r^{2}+a^{2}}dt\wedge dr+d(t+r)\wedge d\varphi ]= \\ 
\qquad =d[\frac{-\gamma _{j}^{\prime \prime }\sqrt{2}}{\pi a}(t+r)(\frac{a}{%
r^{2}+a^{2}}dr+d\varphi )] \\ 
\end{array}
\label{f28}
\end{equation}%
In cartesian coordinates the electric and magnetic fields of the first
solution are 
\begin{equation}
\begin{array}{l}
\overrightarrow{E_{j}^{\prime }}=\frac{\gamma _{j}^{\prime }r^{2}}{%
(r^{4}+a^{2}z^{2})\pi a\sqrt{2}}%
\begin{pmatrix}
\frac{a(ay-rx)}{r^{2}+a^{2}}-\frac{y(r^{2}+a^{2})}{x^{2}+y^{2}} \\ 
\frac{x(r^{2}+a^{2})}{x^{2}+y^{2}}-\frac{a(ax+ry)}{r^{2}+a^{2}} \\ 
-\frac{az}{r}%
\end{pmatrix}
\\ 
\overrightarrow{B_{j}^{\prime }}=\frac{\gamma _{j}^{\prime }r}{%
(r^{4}+a^{2}z^{2})\pi a\sqrt{2}}%
\begin{pmatrix}
\frac{(r^{2}+a^{2})xz}{(x^{2}+y^{2})} \\ 
\frac{(r^{2}+a^{2})yz}{(x^{2}+y^{2})} \\ 
-1%
\end{pmatrix}%
\end{array}
\label{f29}
\end{equation}%
The second solution gives 
\begin{equation}
\begin{array}{l}
\overrightarrow{E_{j}^{\prime \prime }}=\frac{\gamma _{j}^{\prime \prime
}r^{2}\sqrt{2}}{(r^{4}+a^{2}z^{2})\pi a}%
\begin{pmatrix}
\frac{y(r^{2}+a^{2})}{x^{2}+y^{2}}-\frac{a(ay+rx)}{r^{2}+a^{2}} \\ 
\frac{a(ax-ry)}{r^{2}+a^{2}}-\frac{x(r^{2}+a^{2})}{x^{2}+y^{2}} \\ 
-\frac{az}{r}%
\end{pmatrix}
\\ 
\overrightarrow{B_{j}^{\prime \prime }}=\frac{\gamma _{j}^{\prime \prime }r%
\sqrt{2}}{(r^{4}+a^{2}z^{2})\pi a}%
\begin{pmatrix}
\frac{(r^{2}+a^{2})xz}{(x^{2}+y^{2})} \\ 
\frac{(r^{2}+a^{2})yz}{(x^{2}+y^{2})} \\ 
-1%
\end{pmatrix}%
\end{array}
\label{f30}
\end{equation}

The present classical solutions of the null gauge field with colored sources
is expected to lead to confinement, because it has a vortex-like singularity
at $z=\pm r$.

Notice that a non-null solution may exist if $F_{j01}$ and $F_{k\widetilde{0}%
\widetilde{1}}$ do not vanish for $j$ and $k$\ taking values for commuting
generators (of the maximal abelian subgroup). That is, in the interesting
case of $su(3)$\ we have such a distributional solution if $i,j\in \{3,8\}$.
This solution is essentially the abelian solution (\ref{f14}).

\section{PERSPECTIVES}

\setcounter{equation}{0}

The recent experimental results of the LHC experiments at CERN imply that
supersymmetric particles do not exist and subsequently, quantum string
theory does not describe nature. Hence, the 4-dimensional PCFT remains the
only known model, compatible with quantum theory, which provides the general
experimentally observed framework.

The solitonic approach seems to be the computationally convenient procedure.
These techniques have been already developed\cite{MANTON} in the skyrmion
model for the simulation of hadronic dynamics and the
Bogomolny-Prassad-Sommerfield (BPS) model for the experimental search of
monopoles.

In order to facilitate the reader to understand the present possibility to
appear the quark (colored) soliton in the PCFT, I will now recapitulate the
general framework. The LCR-manifolds are special real surfaces of an ambient
complex manifold determined with relations of the form (\ref{i6}). The
ambient complex manifolds of the electron (and its neutrino) are the
intersections of the irreducible (and reducible) static (and stationary)
quadratic surfaces\cite{GRIF} and the lines of $CP^{3}$. That is, they are
submanifolds of the grassmannian manifold $G_{4,2}$. The metric is defined
by a symmetric tensor $g_{\mu \nu }$ determined by the two real and the
complex cotangent vectors of the LCR-structure. The spacetime is of Petrov
type D, while the II and I Petrov types are expected to correspond to the
unstable ($\mu ,\nu _{\mu }$) and ($\tau ,\nu _{\tau }$) leptonic families.
That is, the electron LCR-manifold coincides with the Kerr-Newman
LCR-structure and vanishing gauge field $A_{j\mu }=0$. It happens that its
corresponding (non-null) antisymmetric self-dual tensor defines a closed
2-form, identified with the electromagnetic field with source the electron,
located at the singularity ring of the quadratic submanifold of $G_{4,2}$
(the points which correspond to tangent lines of the hypersurface). The
conserved charge breaks the tetrad-Weyl symmetry down to the ordinary Weyl
symmetry. The conservation of the 4-momentum and the angular momentum in the
linearized Einstein gravity approximation breaks the Weyl-symmetry down to
the Poincar\'{e} symmetry. This picture describes the leptonic sector\cite%
{RAG2018a} where the (gluon) gauge field vanishes. In the present work I
solved the linearized stability PDEs of the static soliton and the gauge
field equations of the model. In both cases solutions have been found.

The essential difference between the conventional solitons and the present
one is at the finite energy requirement. The present PCFT contains gravity
and the energy, momentum and angular momentum are conventionally defined
through the corresponding conserved quantities of the linearized gravity. We
do not need any finiteness condition. On the other hand the pseudo-conformal
transformations\cite{RAG2017} contain the energy-momentum implying conserved
currents. The interesting question is "what is the relation of the
energy-momentum yielded by the corresponding pseudo-conformal transformation
and the conventional one of the linearized Einstein gravity?"

This question is better understood if the action of the model is transcribed
using the LCR-structure coordinates $(z^{\alpha }(x),\;z^{\widetilde{\alpha }%
}(x))$,\ $\alpha =0,\ 1$ implied by the application of the (holomorphic)
Frobenius theorem and defined in (\ref{f16}) are used\cite{RAG2017} instead
of the tangent vectors ($\ell ,m;n,\overline{m}$) 
\begin{equation}
\begin{array}{l}
I_{G}=\int d^{4}x\left[ \det (\partial _{\lambda }z^{a})\ \left\{ (\partial
_{0}x^{\mu })(\partial _{1}x^{\nu })F_{j\mu \nu }\right\} \{(\partial _{%
\widetilde{0}}x^{\rho })(\partial _{\widetilde{1}}x^{\sigma })F_{j\rho
\sigma }\}+c.\ c.\right] \\ 
\\ 
I_{C}=\int d^{4}x\ \epsilon ^{\mu \nu \rho \sigma }[\phi _{0}(\partial _{\mu
}z^{0})(\partial _{\nu }z^{1})(\partial _{\rho }\overline{z^{0}})(\partial
_{\sigma }\overline{z^{1}})+\phi _{\widetilde{0}}(\partial _{\mu }z^{%
\widetilde{0}})(\partial _{\nu }z^{\widetilde{1}})(\partial _{\rho }%
\overline{z^{\widetilde{0}}})(\partial _{\sigma }\overline{z^{\widetilde{1}}}%
)+ \\ 
\qquad \qquad +\phi (\partial _{\mu }\overline{z^{0}})(\partial _{\nu }%
\overline{z^{1}})(\partial _{\rho }z^{\widetilde{0}})(\partial _{\sigma }z^{%
\widetilde{1}})+\overline{\phi }(\partial _{\mu }z^{0})(\partial _{\nu
}z^{1})(\partial _{\rho }\overline{z^{\widetilde{0}}})(\partial _{\sigma }%
\overline{z^{\widetilde{1}}})] \\ 
\end{array}
\label{p2}
\end{equation}%
where the $4\times 4$ matrix ($\partial _{b}x^{\mu }$) is the inverse of ($%
\partial _{\mu }z^{b}$).

This action is explicitly invariant under the following two infinitesimal
pseudo-conformal (LCR-structure preserving) transformations%
\begin{equation}
\begin{array}{l}
\delta z^{\beta }\simeq \varepsilon \psi ^{\beta }(z^{\gamma })\quad ,\quad
\delta z^{\widetilde{\beta }}\simeq \widetilde{\varepsilon }\psi ^{%
\widetilde{\beta }}(z^{\widetilde{\gamma }}) \\ 
\\ 
\delta \phi _{0}=-\phi _{0}[(\partial _{\alpha }\psi ^{\alpha })\varepsilon
+(\overline{\partial _{\alpha }\psi ^{\alpha }})\overline{\varepsilon }] \\ 
\\ 
\delta \phi _{\widetilde{0}}=-\phi _{\widetilde{0}}[(\partial _{\widetilde{%
\alpha }}\psi ^{\widetilde{\alpha }})\widetilde{\varepsilon }+(\overline{%
\partial _{\widetilde{\alpha }}\psi ^{\widetilde{\alpha }}})\overline{%
\widetilde{\varepsilon }}] \\ 
\\ 
\delta \phi =-\phi \lbrack (\partial _{\alpha }\psi ^{\alpha })\varepsilon +(%
\overline{\partial _{\widetilde{\alpha }}\psi ^{\widetilde{\alpha }}})%
\overline{\widetilde{\varepsilon }}]%
\end{array}
\label{p3}
\end{equation}%
Notice that the transformations of the "left" $z^{\alpha }(x)$ and "right" $%
z^{\widetilde{\alpha }}(x)$ structure coordinates are independent, like the
conformal transformations in the ordinary 2-dimensional CFT. The following
"left" and "right" LCR-currents%
\begin{equation}
\begin{array}{l}
J^{\lambda }\equiv -\det (\partial _{\tau }z^{a})\ F_{j01}\psi ^{\gamma
}F_{j\gamma \widetilde{\alpha }}\epsilon ^{\widetilde{\alpha }\widetilde{%
\beta }}(\partial _{\widetilde{\beta }}x^{\lambda })- \\ 
\quad -\epsilon _{\alpha \beta }\psi ^{\alpha }\epsilon ^{\lambda \nu \rho
\sigma }(\partial _{\nu }z^{\beta })[\phi _{0}(\partial _{\rho }\overline{%
z^{0}})(\partial _{\sigma }\overline{z^{1}})+\overline{\phi }(\partial
_{\rho }\overline{z^{\widetilde{0}}})(\partial _{\sigma }\overline{z^{%
\widetilde{1}}})] \\ 
\\ 
\widetilde{J}^{\lambda }\equiv -\det (\partial _{\tau }z^{a})\ \psi ^{%
\widetilde{\gamma }}\epsilon ^{\alpha \beta }F_{j\alpha \widetilde{\gamma }%
}(\partial _{\beta }x^{\lambda })F_{j\widetilde{0}\widetilde{1}}- \\ 
\quad -\epsilon _{\widetilde{\alpha }\widetilde{\beta }}\psi ^{\widetilde{%
\alpha }}\epsilon ^{\lambda \nu \rho \sigma }(\partial _{\nu }z^{\widetilde{%
\beta }})[\phi _{\widetilde{0}}(\partial _{\rho }\overline{z^{\widetilde{0}}}%
)(\partial _{\sigma }\overline{z^{\widetilde{1}}})+\phi (\partial _{\rho }%
\overline{z^{0}})(\partial _{\sigma }\overline{z^{1}})] \\ 
\end{array}
\label{p4}
\end{equation}%
are derived.

The independent conserved quantities are implied by the formally independent
functions%
\begin{equation}
\begin{array}{l}
\psi ^{\alpha }(z^{\gamma })=-(z^{\alpha
})(z^{0})^{m_{0}}(z^{1})^{m_{1}}\quad \Longrightarrow \quad T_{%
\overrightarrow{m}}^{(\alpha )}=\int d^{4}xJ_{(\alpha )\overrightarrow{m}%
}^{0} \\ 
\\ 
\psi ^{\widetilde{\beta }}(z^{\widetilde{\gamma }})=-(z^{\widetilde{\beta }%
})(z^{\widetilde{0}})^{m_{0}}(z^{\widetilde{1}})^{m_{1}}\quad
\Longrightarrow \quad T_{\overrightarrow{m}}^{(\widetilde{\beta })}=\int
d^{4}xJ_{(\widetilde{\beta })\overrightarrow{m}}^{0} \\ 
\end{array}
\label{p5}
\end{equation}%
The implied "left" and "right" LCR-currents determine the following
independent conserved quantities%
\begin{equation}
\begin{array}{l}
\psi ^{\alpha }(z^{\gamma })=-(z^{\alpha
})(z^{0})^{m_{0}}(z^{1})^{m_{1}}\quad \Longrightarrow \quad T_{%
\overrightarrow{m}}^{(\alpha )}=\int d^{4}xJ_{(\alpha )\overrightarrow{m}%
}^{0} \\ 
\\ 
\psi ^{\widetilde{\beta }}(z^{\widetilde{\gamma }})=-(z^{\widetilde{\beta }%
})(z^{\widetilde{0}})^{m_{0}}(z^{\widetilde{1}})^{m_{1}}\quad
\Longrightarrow \quad T_{\overrightarrow{m}}^{(\widetilde{\beta })}=\int
d^{4}xJ_{(\widetilde{\beta })\overrightarrow{m}}^{0} \\ 
\end{array}
\label{p7}
\end{equation}%
where $\overrightarrow{m}=(m_{0},m_{1})$ and $m_{\beta }$\ take integer
values. Do not confuse this integer symbol $m_{\beta }$\ with the
corresponding tetrad vector.

Considering explicit examples, one can easily see that the currents contain
the generators of the Poincar\'{e} group, like the 2-d analogue CFT. This
formulation and the solitonic procedure could (in principle) permit us to
compute the Poincar\'{e} charges and relate them with those conventionally
found from linearized Einstein gravity.

\end{document}